\documentclass{mn}

\usepackage{amsmath}
\usepackage{amssymb}
\usepackage{subfigure}
\usepackage{graphicx}

 \font\sevenrm=cmr7 scaled 1000
\def\gsim{\;\lower4pt\hbox{${\buildrel\displaystyle >\over\sim}$}\;}
\def\lsim{\;\lower4pt\hbox{${\buildrel\displaystyle <\over\sim}$}\;}
\def\grls{\;\lower4pt\hbox{${\buildrel\displaystyle >\over <}$}\;}
\begin{document}
\title[Similarity Shock Solutions]
{Spherical Isothermal Self-Similar Shock Flows }
\author[F.-Y. Bian and Y.-Q. Lou]
       {Fu-Yan Bian$^1$ and
       Yu-Qing Lou$^{1,2,3,4}$ \\
        $^1$Physics Department and Tsinghua Center for
Astrophysics (THCA), Tsinghua University, Beijing 100084, China;\\
        $^2$Centre de Physique des Particules de Marseille
(CPPM)/Centre National de la Recherche Scientifique (CNRS)\\
\ \ \qquad
/Institut National de Physique Nucl\'eaire et de
Physique des Particules (IN2P3) et Universit\'e\\ \ \ \qquad
de la M\'editerran\'ee Aix-Marseille II, 163,
Avenue de Luminy Case 902 F-13288 Marseille, Cedex 09, France;\\
        $^3$Department of Astronomy and Astrophysics, The University
        of Chicago, 5640 S Ellis Ave, Chicago, IL 60637 USA;\\
        $^4$National Astronomical Observatories, Chinese Academy
        of Sciences, A20, Datun Road, Beijing, 100012 China.\\
      E-mail: \ \ louyq@mail.tsinghua.edu.cn\ \ \ lou@oddjob.uchicago.edu}
\date{Accepted 2005 ............;
      Received 2005.............;
      in original form 2005......}
\pagerange{\pageref{firstpage}--\pageref{lastpage}} \pubyear{2005}

\maketitle

\label{firstpage}

\begin{abstract}
We explore self-similar dynamical processes in a spherical
isothermal self-gravitational fluid with an emphasis on shocks and
outline astrophysical applications of such shock solutions. The
previous similarity shock solutions of Tsai \& Hsu and of Shu et al.
may be classified into two types: Class I solutions with downstream
being free-fall collapses and Class II solutions with downstream
being Larson-Penston (LP) type solutions. By the analyses of Lou
\& Shen and Shen \& Lou, we further construct similarity shock
solutions in the `semi-complete space'.
These general shock solutions can accommodate and model dynamical
processes of radial outflows (wind), inflows (accretion or
contraction), subsonic oscillations, and free-fall core collapses
all with shocks in various settings such as star-forming molecular
clouds, `champagne flows' in H{\sevenrm II} regions around luminous
massive OB stars or surrounding quasars, dynamical connection between
the asymptotic giant branch phase to the proto-planetary nebula phase
with a central hot white dwarf as well as accretion shocks around
compact objects such as white dwarfs, neutron stars, and supermassive
black holes. By a systematic exploration, we are able to construct
families of infinitely many discrete Class I and Class II solutions
matching asymptotically with a static outer envelope of singular
isothermal sphere; the shock solutions of Tsai \& Hsu form special
subsets. These similarity shocks travel at either subsonic or
supersonic constant speeds. We also construct twin shocks as well
as an `isothermal shock' separating two fluid regions of two
different yet constant temperatures.
\end{abstract}
\begin{keywords}
hydrodynamics --- stars: AGB and post AGB --- stars: formation
--- planetary nebulae: general --- H {\sevenrm II} regions
--- quasars: general
\end{keywords}

\section{Introduction}

Self-gravitational inflows (contractions or accretions or
collapses) and outflows (expansions or winds) of an isothermal
fluid with spherical symmetry have been investigated over the
past several decades in various astrophysical contexts (e.g.,
star formations, supernova explosions, formation and evolution
of galaxy clusters etc). When a fluid system is sufficiently
away from initial and boundary conditions, it may evolve into
self-similar behaviours (e.g., Sedov 1959; Landau \& Lifshitz 1959).
Under isothermal condition and spherical symmetry, Larson (1969a,b)
and Penston (1969a,b) independently found self-similar solutions,
now commonly referred to as Larson-Penston (LP) type solutions.
Shu (1977) obtained another class of similarity solutions containing
the so-called `expansion wave collapse solution' (EWCS) that describes
an expanding region of core collapse from a surrounding molecular cloud.
In addition to these solutions, Hunter (1977) derived a class of
discrete similarity solutions in the `complete solution space' from
$t\rightarrow-\infty$ to $t\rightarrow+\infty$. These solutions may
represent clouds which are accreted to a central mass point as an inward
propagating compression wave driven by external gas pressure. Whitworth
\& Summers (1985) expanded the solution space by introducing a bounded
two-parameter continuum \{$z_0$, $w_0$\} to the known discrete solutions.
The two parameters $z_0$ and $w_0$ correspond to $\alpha_0$ and $m_0$ in
the solutions of Shu (1977), which characterize the asymptotic solution
forms as $t\rightarrow-\infty$ and $t\rightarrow +\infty$ respectively.
Hunter (1986) noted that the additional bands of solutions of Whitworth
\& Summers (1985) involve weak discontinuities across the sonic critical
line (Boily \& Lynden-Bell 1995) and may be unstable to perturbations
(e.g., Lazarus 1981).
Generalizing these earlier results, Lou \& Shen (2004) recently
obtained a new class of self-similar isothermal solutions, referred to
as `envelope expansion with core collapse' (EECC) solutions, which are
characterized by an interior free-fall collapse towards the core with
an exterior envelope expanding at a constant radial speed at large
radii. A broad class of EECC solutions without crossing the sonic
critical line can be readily obtained; meanwhile, a special class of
infinitely many discrete EECC solutions crossing the sonic critical
line twice analytically can also be constructed. Boily \& Lynden-Bell
(1995) and Murakami et al. (2004) studied self-similar solutions for
spherical collapses with radiative effects included. The work
of Fatuzzo et al. (2004) treated protostellar collapse in a polytropic
gaseous sphere and found self-similar solutions with nonzero inward
speed at large radii (see earlier emphasis of Shen \& Lou 2004 on this
point); in their model consideration, the infall rates are determined
by the inward speed $v_{\infty}$, the overdensity $\Lambda$, the
polytropic index $\Gamma$ for the background equation of state, and
the polytropic index $\gamma$ for the dynamic equation of state; they
did not explore polytropic solutions crossing the sonic critical line.

Based on prior results, Tsai \& Hsu (1995) found two classes of
isothermal self-similar shock solutions referred to as Class I
and II solutions which connects a static singular isothermal
sphere (SIS) envelope to a free-fall solution and a LP solution
respectively (see \S\ 3.1.1).
Shu et al. (2002) expanded Tsai \& Hsu's LP similarity shock
solution to model the so-called `champagne flows' driven by a
shock into H{\sevenrm II} regions surrounding luminous massive
OB stars (see \S\ 3.1.2).
Further extending the work of Tsai \& Hsu (1995) and Shu et al.
(2002), Shen \& Lou (2004) introduced Class I shock solutions
for dynamical evolution of young stellar objects
and Class II solutions to model `champagne flows' of H{\sevenrm II}
regions surrounding OB stars. More generally, they pointed out that
shock similarity solutions can be matched with various types of
upstream solutions (see \S\ 3.1.3).

In this paper, we construct self-similar isothermal
shock solutions by exploring diverse possibilities.
In \S\ 2, we summarize the basic isothermal fluid equations,
the self-similar transformation and the isothermal shock
conditions. In reference to previous similarity shock solutions,
we derive in \S\ 3 an infinite number of discrete shock solutions
for Class I and Class II categories by exploring the $\alpha-v$
phase diagram and present new similarity shock solutions: twin
shock solution and two-temperature flows separated by shocks.
In \S\ 4, we discuss astrophysical applications of these shock
solutions. In \S\ 5, we summarize and discuss our results.

\section{Formulation of Shock Conditions}

We begin with the fluid equations in the Eulerian form in the
spherical polar coordinates $(r,\theta,\phi)$. By spherical
symmetry, the mass conservation is given by
\begin{equation}
\frac{\partial M}{\partial t}+u\frac{\partial M}{\partial r}=0\ ,
\qquad\qquad
\hspace{0.5cm} \frac{\partial M}{\partial r}=4\pi r^2\rho\ ,
\label{eq mass conservation}
\end{equation}
where $M(r,t)$ is the enclosed mass within radius $r$ at
time $t$. The equivalent form of continuity equation
(\ref{eq mass conservation}) is simply
\begin{equation}
\frac{\partial\rho}{\partial t}
+\frac{1}{r^2}\frac{\partial}{\partial r}(r^2\rho u)=0\ ,
\label{eq rewrite mass conservation}
\end{equation}
where $\rho$ is the gas mass density and $u$ is
the radial flow speed.
The radial momentum equation reads
\begin{equation}
\frac{\partial u}{\partial t}+u\frac{\partial u}{\partial r}
=-\frac{a^2}{\rho}\frac{\partial \rho}{\partial r}-\frac{GM}{r^2}\ ,
\label{eq radial momentum equation}
\end{equation}
where $G\equiv6.67\times 10^{-8}$ dyn cm$^2$ g$^{-2}$ is the
gravitational constant, $a\equiv (p/\rho)^{1/2}$ is the isothermal
sound speed and $p$ is the gas pressure. The Poisson equation is
automatically satisfied by equations (\ref{eq mass conservation})
and (\ref{eq radial momentum equation}).

The dimensionless independent similarity variable is
$x=r/(at)$ and the similarity transformations are
\begin{eqnarray}
\qquad \rho(r,t)=\frac{\alpha(x)}{4\pi Gt^2}\ ,&&\hspace{0.3cm}
M(r,t)=\frac{a^3t}{G}m(x)\ ,
\nonumber \\
\qquad u(r,t)=av(x)\ ,&&\hspace{0.3cm} \Phi(r, t)=a^2\phi\ ,
\label{eq reduced parameter}
\end{eqnarray}
where $\alpha(x)$, $m(x)$ and $v(x)$ are dimensionless reduced
variables for mass density $\rho(r,t)$, enclosed mass $M(r,t)$
and radial flow speed $u(r,t)$, respectively; $\Phi$ is the
gravitational potential such that $-\partial\Phi/\partial r
=-GM(r,t)/r^2$ and $\phi(x)$ is its reduced form. These reduced
variables depend only on $x$ (e.g., Shu 1977; Lou \& Shen 2004).

Transformation (\ref{eq reduced parameter}) and
equation (\ref{eq mass conservation}) lead to
\begin{equation}
m+(v-x)\frac{\hbox{d}m}{\hbox{d}x}=0\ ,
\qquad\hspace{0.5cm}\frac{\hbox{d}m}{\hbox{d}x}=x^2\alpha\ ,
\end{equation}
which gives $m(x)$ in terms of $x$, $v$ and $\alpha$, namely
\begin{equation}
m(x)=x^2\alpha(x-v)\ .
\label{mx}
\end{equation}
The physical requirement of a positive $m(x)$
corresponds to the solution regime of $x-v>0$.

By transformation (\ref{eq reduced parameter}) and solution
(\ref{mx}) in equations (\ref{eq rewrite mass conservation})
and (\ref{eq radial momentum equation}), we obtain a pair of
coupled nonlinear ordinary differential equations (ODEs):
\begin{equation}
[(x-v)^2-1]\frac{\hbox{d}v}{\hbox{d}x}
=\left[\alpha(x-v)-\frac{2}{x}\right](x-v)\ ,
\label{eq ODE1}
\end{equation}
\begin{equation}
[(x-v)^2-1]\frac{1}{\alpha}\frac{\hbox{d}\alpha}{\hbox{d}x}
=\left[\alpha-\frac{2}{x}(x-v)\right](x-v)\ ,
\label{eq ODE2}
\end{equation}
[see eqs (11) and (12) of Shu (1977) and eqs (9) and (10) in
Lou \& Shen (2004)]. In the two nonlinear ODEs, the critical
line is characterized by $(x-v)^2-1=0$ and $\alpha=2/x$.


The analytical asymptotic solutions of ODEs (\ref{eq ODE1})
and (\ref{eq ODE2}) are shown below (see Lou \& Shen 2004
for details).

(i) For $x\rightarrow +\infty$, the leading order terms are
\begin{eqnarray}
&&v\rightarrow V\ ,
\qquad\hspace{0.5cm} \alpha\rightarrow \frac{A}{x^2}\ ,
\qquad\hspace{0.5cm} m\rightarrow Ax\ ,
\label{eq asymptotic solution x-infty}
\end{eqnarray}
where $V$ and $A$ are two independent constants, referred to
as velocity and mass parameters, respectively. The envelope
expansion approaches a constant speed $V$ at large radii.

(ii) For $x\rightarrow 0^{+}$, the leading terms are
\begin{eqnarray}
&&v\rightarrow -\left(\frac{2m_0}{x}\right)^{1/2}\ ,
\hspace{0.2cm}
\alpha\rightarrow
\left(\frac{m_0}{2x^3}\right)^{1/2}\ ,
\hspace{0.2cm} m\rightarrow m_0
\label{eq free-fall}
\end{eqnarray}
describing a central free-fall collapse with a constant
dimensionless parameter $m_0$ for central mass accretion
rate (Shu 1977; Lou \& Shen 2004; Shen \& Lou 2004; Yu
\& Lou 2005).

For the central LP-type solutions as $x\rightarrow 0^{+}$,
the leading order terms are
\begin{eqnarray}
&&v\rightarrow \frac{2}{3}x\ ,
\qquad\hspace{0.2cm} \alpha\rightarrow B\ ,
\qquad\hspace{0.2cm} m\rightarrow Bx^3/3
\label{eq LP solution}
\end{eqnarray}
with a finite reduced central mass density $B$.

(iii) Along the sonic critical line, the
two eigensolutions are respectively
\begin{eqnarray}
&&-v(x)=(1-x_{\ast})+\left(\frac{1}{x_{\ast}}
-1\right)(x-x_{\ast})+\cdots\ ,
\nonumber \\
&&\alpha(x)=\frac{2}{x_{\ast}}-\frac{2}{x_{\ast}}
\left(\frac{3}{x_{\ast}}-1\right)(x-x_{\ast})+\cdots
\end{eqnarray}
\label{eq critical solution (1)}
for type 1 solutions, and
\begin{eqnarray}
&&-v(x)=(1-x_{\ast})-\frac{1}{x_{\ast}}(x-x_{\ast})+\cdots\ ,
\nonumber\\
&&\alpha(x)=\frac{2}{x_{\ast}}
-\frac{2}{x_{\ast}^2}(x-x_{\ast})+\cdots
\label{eq critical solution (2)}
\end{eqnarray}
for type 2 solutions, as defined by Shu (1977) and used in the
subsequent work [e.g. Hunter (1977); Whitworth \& Summers (1985);
Lou \& Shen (2004); Shen \& Lou (2004)]. These analytical asymptotic
solutions provide necessary starting conditions for a fourth-order
Runge-Kutta integration scheme (Press et al. 1986) to obtain global
solutions numerically. We derive various classes of numerical
solutions from the two coupled nonlinear ODEs (\ref{eq ODE1})
and (\ref{eq ODE2}) globally.

For an isothermal gas, cooling effects by dust grains allow efficient
radiation of compressional heat generated during a collapse. The energy
conservation involves radiative losses (e.g., Courant \& Friedrichs
1976; Spitzer 1978). Under the isothermal approximation, we need to
consider conservations of mass and momentum across a shock, namely
\begin{equation}
\rho_d(u_d-u_s)=\rho_u(u_u-u_s)\ ,
\label{eq shock mass
conservation}
\end{equation}
\begin{equation}
{a_d}^2\rho_d+\rho_du_d(u_d-u_s)=a_u^2\rho_u+\rho_uu_u(u_u-u_s)\ ,
\label{eq shock momentum conservation}
\end{equation}
where $u$, $a$ and $\rho$ are the radial gas flow speed, the
isothermal sound speed and the mass density, respectively, and
subscripts d (u) denote the downstream (upstream) of a shock,
$u_s=a_dx_{sd}=a_ux_{su}=r_s/t$ is the outward moving speed of
the shock with $r_s$ being the shock radius. Using dimensionless
reduced variables $v(x)$, $x$, and $\alpha(x)$ to replace $u(r,t)$
and $\rho(r,t)$, we derive the isothermal shock jump conditions in
terms of $v(x)$, $x$, and $\alpha(x)$, namely
\begin{equation}
\alpha_d(v_d-x_{sd})a_d=\alpha_u(v_u-x_{su})a_u\ ,
\label{eq shock redeuced mass conservation}
\end{equation}
\begin{equation}
{a_d}^2\alpha_d+{a_d}^2\alpha_dv_d(v_d-x_{sd})
={a_u}^2\alpha_u+{a_u}^2\alpha_uv_u(v_u-x_{su})\ .
\label{eq shock redeuced momentum conservation}
\end{equation}
For an isothermal sound speed ratio $\tau=a_d/a_u$
and $\tau x_{sd}=x_{su}$, we derive
\begin{equation}
v_d-x_{sd}-\tau(v_u-x_{su})=(\tau
v_d-v_u)(v_u-x_{su})(v_d-x_{sd})\ ,
\label{eq shock condition (1) for free tau}
\end{equation}
\begin{equation}
\alpha_d/\alpha_u=(v_u-x_{su})/[\tau(v_d-x_{sd})]\ \label{eq shock
condition (2) for free tau}
\end{equation}
(Shen \& Lou 2004).
For $\tau=1$, eqns
(\ref{eq shock condition (1) for free tau}) and
(\ref{eq shock condition (2) for free tau}) become
\begin{equation}
(v_u-x_s)(v_d-x_s)=1
\label{eq shock condition (1) for tau=1}
\end{equation}
\begin{equation}
\alpha_d/\alpha_u=(v_u-x_s)/(v_d-x_s)\ ,
\label{eq shock condition (2) for tau=1}
\end{equation}
where $x_s=x_{su}=x_{sd}$ is the reduced shock speed.

Our main goal is to solve coupled nonlinear ODEs (\ref{eq ODE1})
and (\ref{eq ODE2}) and to apply the jump conditions across
various shocks to match with reasonable boundary conditions for
constructing global similarity shock solutions. Conceptually,
this would help to conceive various possible scenarios involving
shocks. With proper adaptations, this would lead to interesting
physical models for various astrophysical systems.

\section{Self-Similar Shock Solutions}

We can construct various similarity solutions from nonlinear ODEs
(\ref{eq ODE1}) and (\ref{eq ODE2}) with shocks satisfying shock
jump conditions (\ref{eq shock condition (1) for free tau}) and
(\ref{eq shock condition (2) for free tau}) to match with different
analytical asymptotic solutions. In general, these similarity
solutions can describe radial outflows (expansions or winds),
inflows (contractions or accretions), oscillations, free-fall
collapses with one or more shocks. We explore the range of the
independent similarity variable $x$ in the `semi-complete space',
i.e., from $x\rightarrow 0^{+}$ to $x\rightarrow +\infty$. By the
invariant property for the time reversal of the equations, it is
easy to establish the correspondence between solutions in the
`complete solution space' (Hunter 1977) and solutions in the
`semi-complete solution space' (Lou \& Shen 2004).

In this section, we first review earlier results of Tsai \& Hsu
(1995), Shu et al. (2002) and Shen \& Lou (2004). We then derive
two classes of infinitely many discrete shock solutions including
the twin shock solutions. Finally, we study shock solutions with
the shock separating the flows of two different yet constant
temperatures.

\subsection{Previous Shock Solution Results}

We now review the similarity solution structure in reference to
earlier results of Tsai \& Hsu (1995), Shu et al. (2002), and Shen
\& Lou (2004). The common feature of these shock solutions is that
they have been constructed with $\tau=1$ and contain only one shock.
As $x\rightarrow 0^{+}$, these shock solutions can match with inner
free-fall collapses across the sonic critical line or with inner
LP solutions, while as $x\rightarrow+\infty$, they match with the
analytical asymptotic solution (i) by equation
(\ref{eq asymptotic solution x-infty}). These solutions can be
broadly classified by their behaviours near $x\rightarrow 0^{+}$
into two types: Class I whose downstream solutions are free-fall
collapses and Class II whose downstream solutions are LP-type
solutions.

\subsubsection{Shock Solutions of Tsai \& Hsu (1995) }
\begin{figure}
\begin{center}
\includegraphics[scale=0.43]{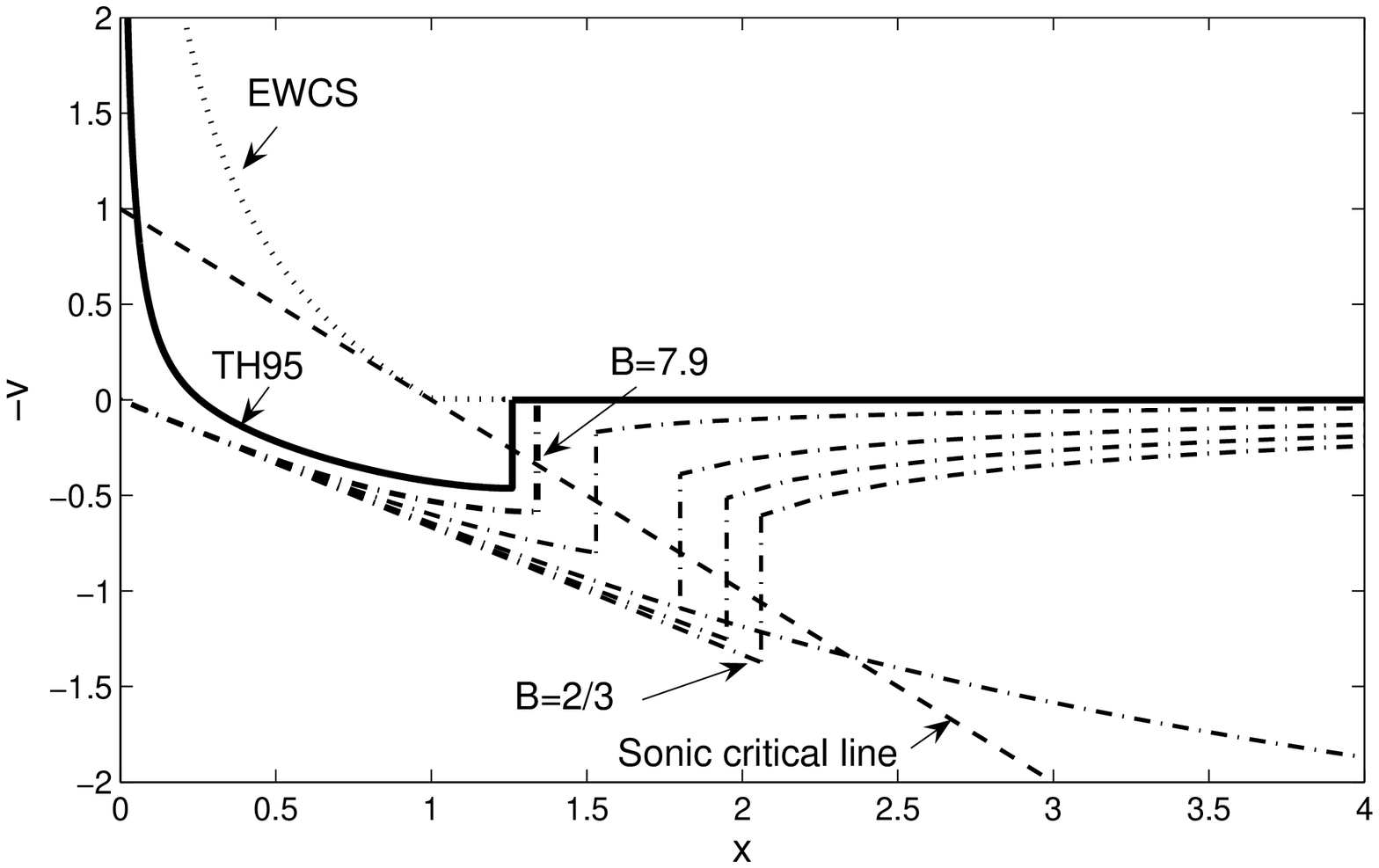}%
\hspace{1in}%
\includegraphics[scale=0.43]{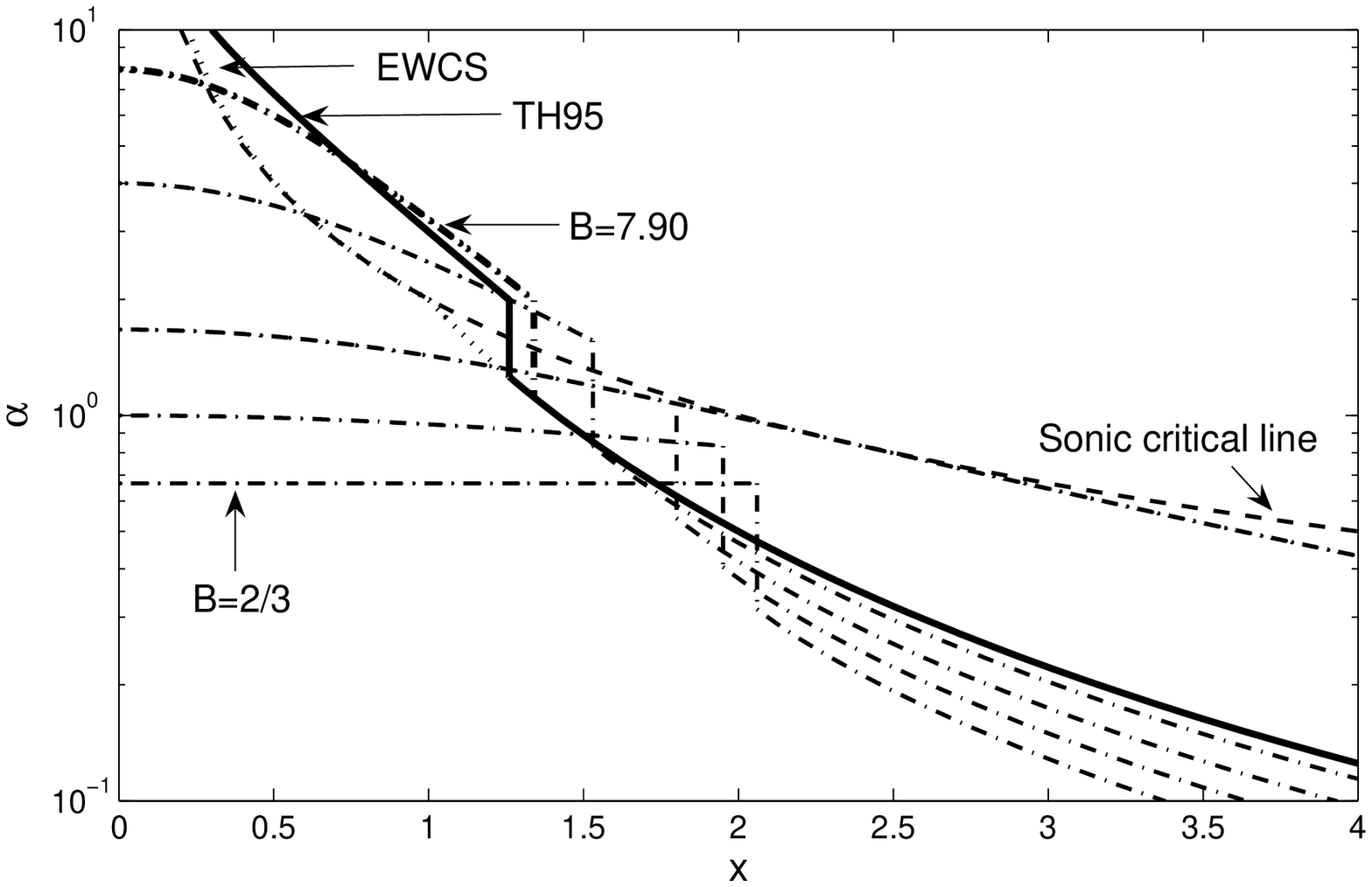}
\caption{The reduced radial speed $-v(x)$ in linear scale (upper
panel) and the reduced mass density $\alpha(x)$ in logarithmic
scale (lower panel) versus $x$ in linear scale; the dashed curves
correspond to the sonic critical line, the dotted curves
correspond to the EWCS of Shu (1977), the heavy solid curve and
the heavy dash-dotted curve of Tsai \& Hsu (1995) correspond to
the shock solutions matching with the inner free-fall collapses
and with inner LP-type solutions to a static SIS envelope
respectively, the inner free-fall solution crosses the sonic
critical line at $x_{\ast}=0.0554$ and the $B$ parameter of the
inner LP solution is $B=7.90$; the light dash-dotted curves are
solutions with asymptotic breezes at large $x$ found by Shu et al.
(2002) with the different curves from top to bottom corresponding
to the parameter $B=2/3$, 1, 1.67 and 4 respectively in both upper
and lower panels.}
\label{fig Tsai Shu}
\end{center}
\end{figure}


Shu (1977) obtained the EWCS (the dotted curve in Figs.~\ref{fig
Tsai Shu}) with the mass density parameter $A\rightarrow2^{+}$
and the velocity parameter $V=0$ in equation (\ref{eq asymptotic
solution x-infty}), and the solution touches the sonic critical
line at $x_{\ast}=1$ with $m_0=0.975$.
Tsai \& Hsu (1995) considered a self-similar shock travelling
into a static SIS envelope with $\tau=1$, $x_{sd}=x_{su}=x_{s}$,
$v_u=0$ and $\alpha_{u}=2/x_s^2$, such that the shock jump
conditions (\ref{eq shock condition (1) for tau=1}) and
(\ref{eq shock condition (2) for tau=1}) appear in the form of
\begin{equation}
v_d=x_s-\frac{1}{x_s}\ ,\qquad\qquad\hspace{0.3cm}\alpha_d=2\ .\quad
\label{eq shock condition for static envelope }
\end{equation}
They showed two specific examples for the two classes of
similarity shock solutions: Class I solution given by heavy solid
curves in the two panels of Fig.~\ref{fig Tsai Shu} and Class II
solution given by heavy dashed-dotted curves in the two panels of
Fig.~\ref{fig Tsai Shu}. This Class I solution crosses the sonic
critical line at $x_{\ast}=0.0554$ and the shock location is at
$x_{s}=1.26$ for an outgoing shock travelling at a constant speed
of 1.26 times the sound speed $a$. As this Class I solution
connects the inner free-fall collapse through an expansion into
a static SIS envelope by a shock, the solution has the same
properties of the EWCS such that the outer gas envelope remains
static with a $r^{-2}$ density profile at large radii
and the central gas falls towards the core at a constant rate
$m_{0}=0.105$ with a $r^{-3/2}$ density profile and a $r^{-1/2}$
velocity profile. The central mass accretion rate $\dot{M}=m_0a^3/G$
in this similarity shock solution is a factor of $\sim 0.1$ that
predicted by the EWCS of Shu (1977). For the Class II solution,
a central expansion with a finite central density (LP solution)
matches to a static SIS envelope across a shock; it shows a higher
shock speed and strength. The shock location is at $x_{s}=1.34$
for a constant shock expansion speed of $1.34$ times the sound
speed $a$ and a reduced core density $B\cong 7.9$.

\subsubsection{Shock Solutions of Shu et al. (2002)}

%
Shu et al. (2002) extended the Class II solution of Tsai \& Hsu (1995)
and used these solutions to model `champagne flows' of H{\sevenrm II}
regions surrounding massive OB stars. By varying the $B$ parameter
with fixed $V=0$, they obtained a class of outflow solutions and
found both shock speed and strength decreasing with increasing $B$.
In Fig.~\ref{fig Tsai Shu}, we plot different solutions (dash-dotted
curves) with different reduced central density parameter $B=2/3,\ 1,\
1.67,\ 4,\ 7.90$, corresponding to the mass parameter $A=1.09,
\ 1.28,\ 1.52,\ 1.832,\ 2.0$ respectively, and also corresponding to
the shock location $x_{s}=2.06,\ 1.95,\ 1.80,\ 1.53,\ 1.34$,
respectively. As $B\rightarrow 0^{+}$ (e.g., $B=10^{-6}$) numerically,
the mass parameter $A$ also approaches $0^{+}$, and the reduced speed
$v(x)$ converges to an invariant form with an invariant fastest and
strongest shock solution where the shock is located at $x_s=2.56$
(see the heavy solid curve in Fig.~\ref{fig Shen and Lou 2}).
%

\subsubsection{Shock Solutions of Shen \& Lou (2004) }
\begin{figure}
\begin{center}
\includegraphics[scale=0.43]{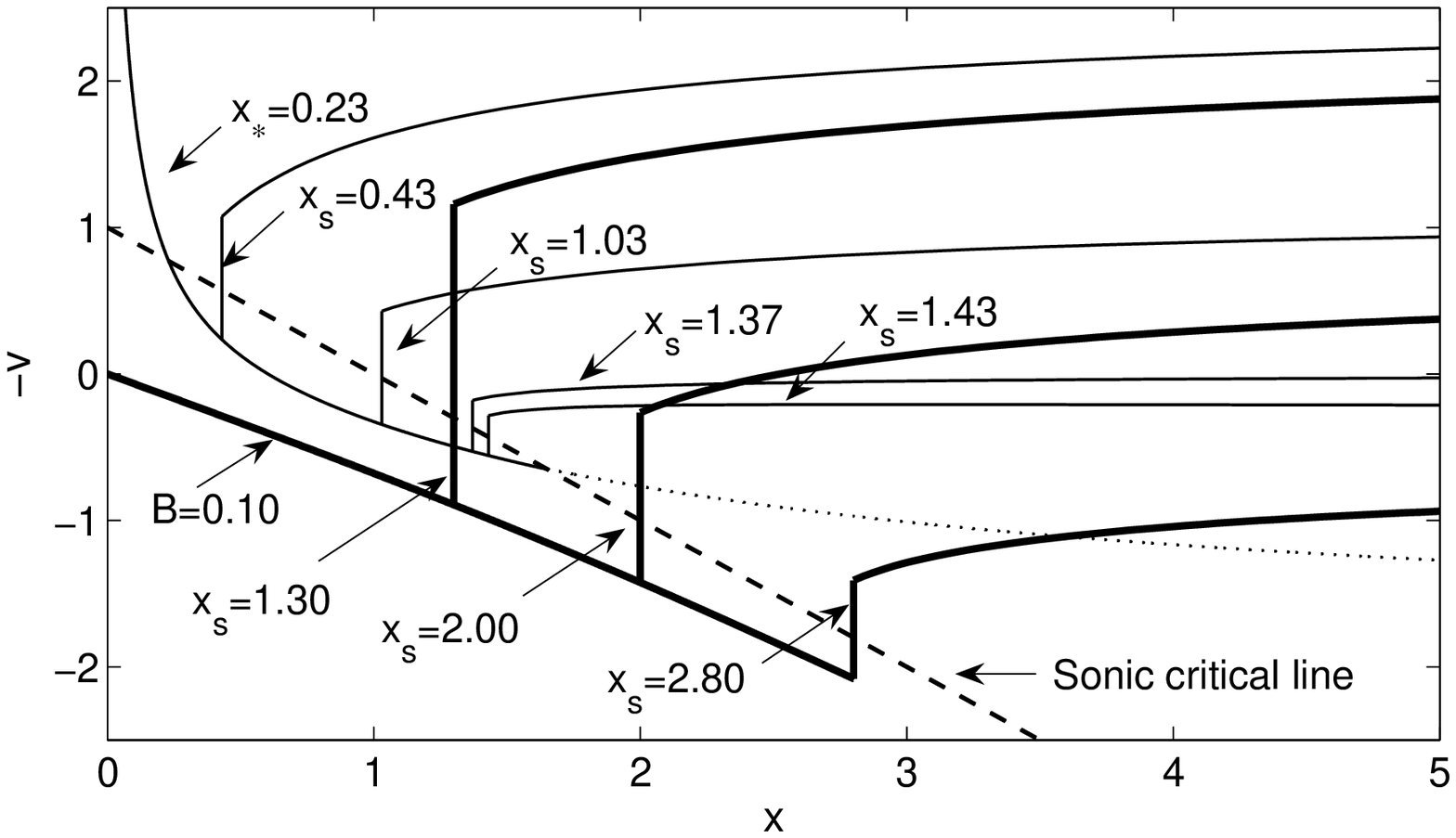}%
\hspace{1in}%
\includegraphics[scale=0.43]{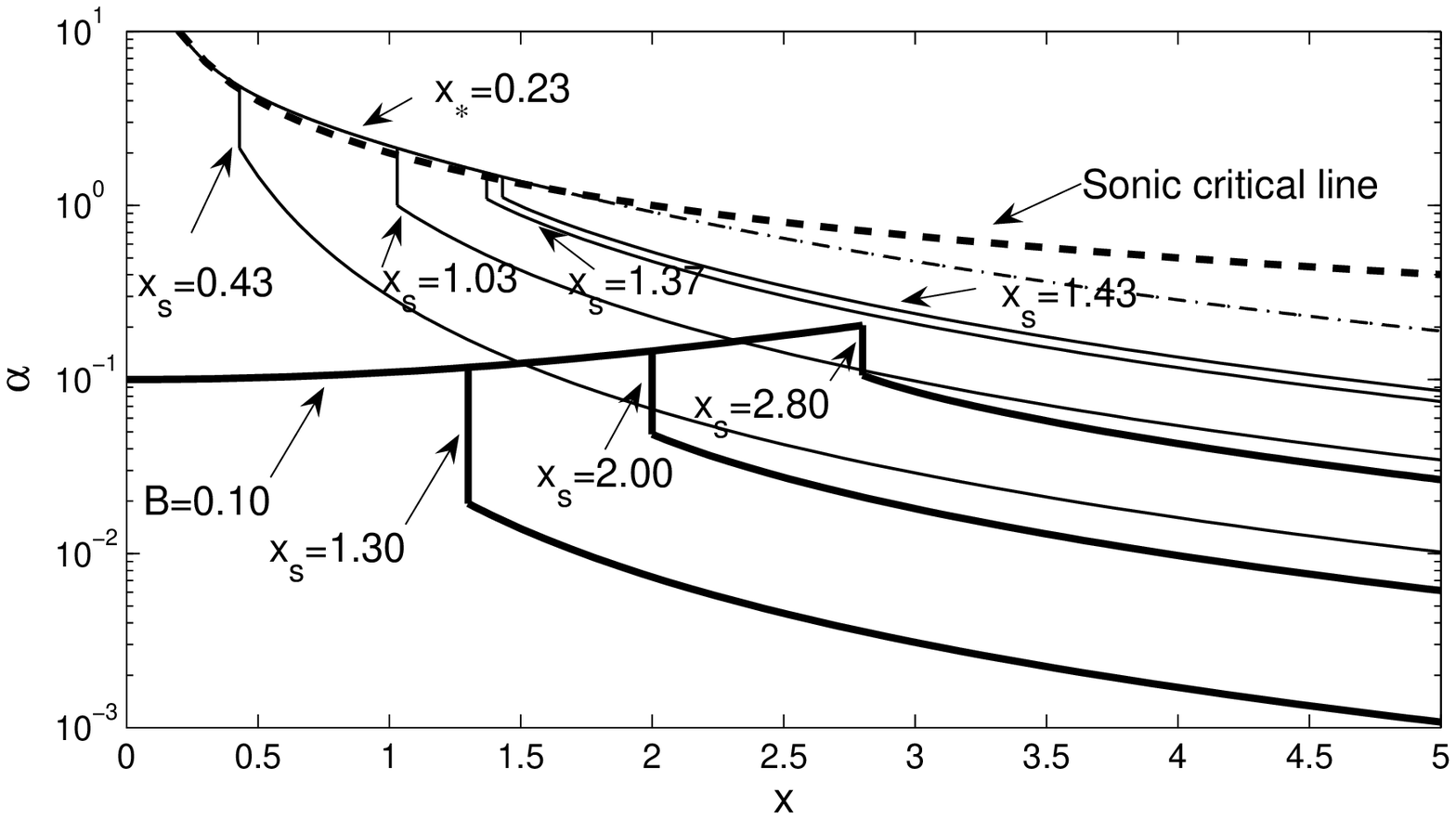}

\caption{The dimensionless negative reduced radial speed $-v(x)$
(top panel) and reduced mass density $\alpha(x)$ (bottom panel)
versus $x$. The dashed lines in both panels are the sonic
critical line; the light solid curves are illustrating examples
of Class I solutions with the downstream solution crossing the
sonic critical line at $x_{\ast}=0.23$ and the shock location
$x_s$ at $0.43,\ 1.03,\ 1.37,\ 1.43$, respectively; the heavy
solid curves are illustrating examples of Class II solutions
with the reduced density of the central core $B=0.1$ and the
shock location $x_s$ at 1.30, 2.00 and 2.80, respectively.}
\label{fig Shen and Lou 1}
\end{center}
\end{figure}
\begin{figure}
\begin{center}
\includegraphics[scale=0.43]{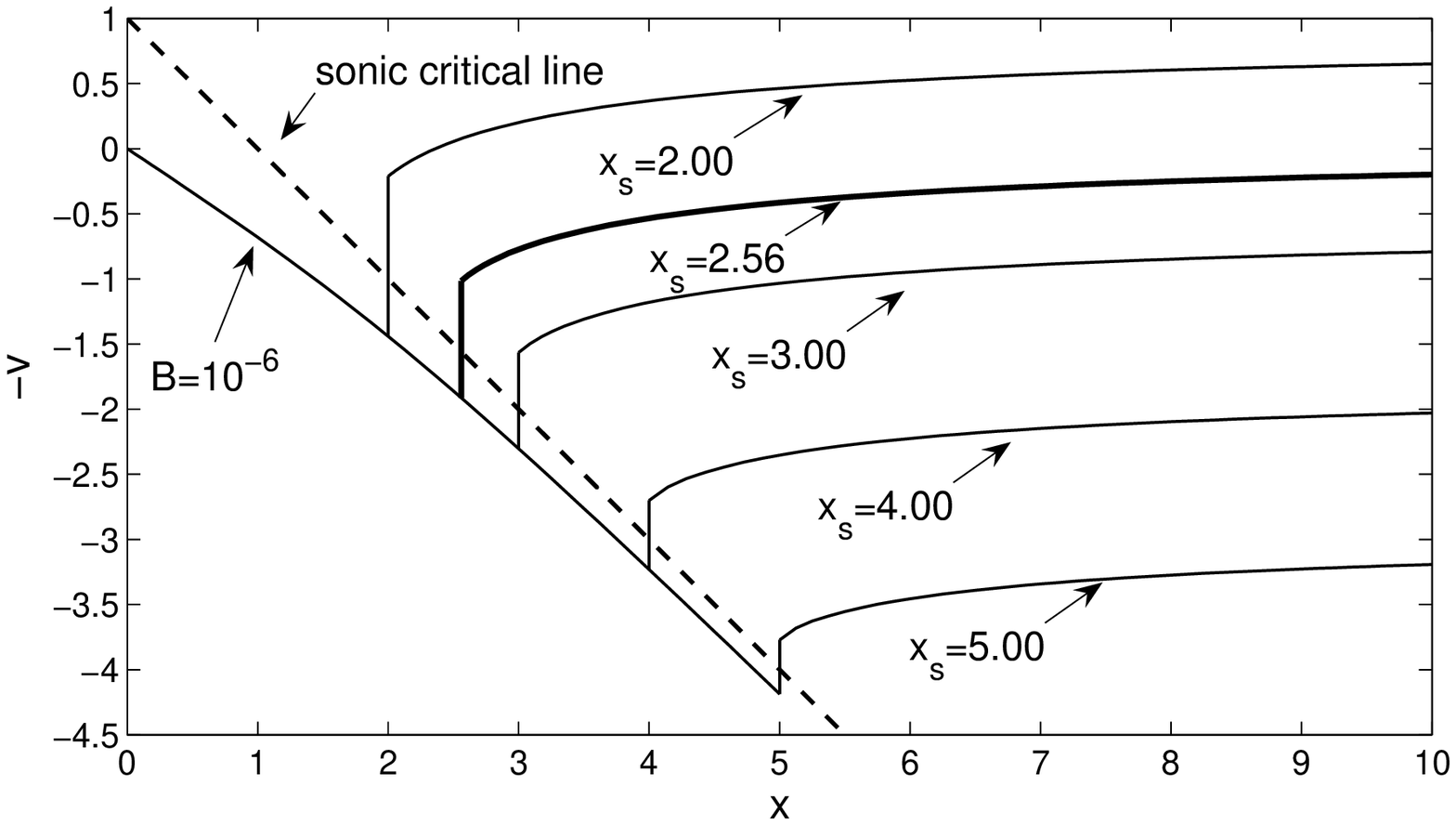}%
\hspace{1in}%
\includegraphics[scale=0.43]{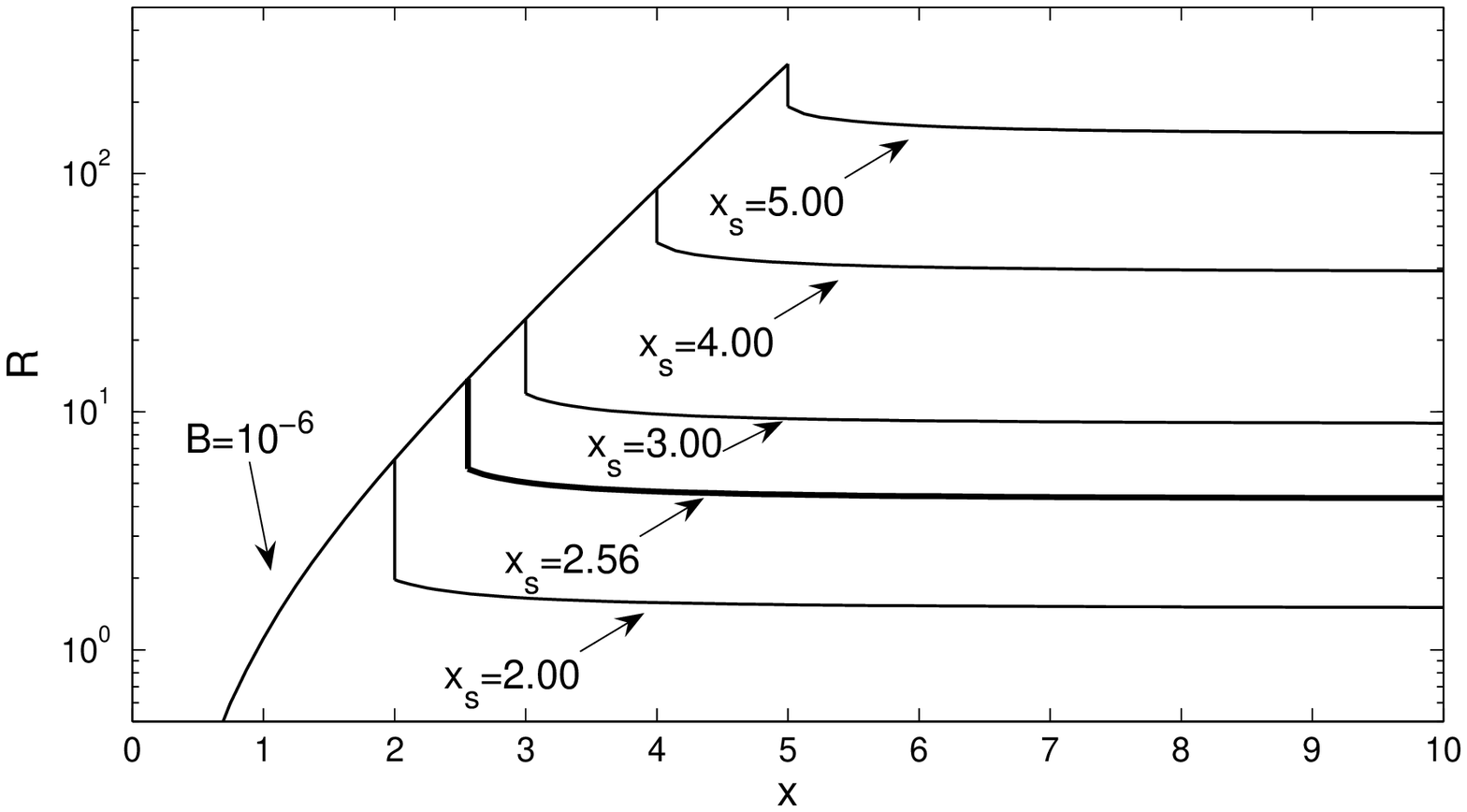}

\caption{The dimensionless reduced speed $v(x)$ (top panel) and a
scaled reduced mass density $R(x)$ where $R(x)\equiv x^2\alpha(x)/B$
(bottom panel) versus $x$. The dashed lines are the sonic critical
line; the solid curves are the family of Class II solutions in the
invariant form with the reduced density of the central core
$B\rightarrow0^{+}$ (viz., $B=10^{-6}$); the heavy solid line is
the so-called `champagne breeze' solution of Shu et al. (2002). }
\label{fig Shen and Lou 2}
\end{center}
\end{figure}

By constructing a new family of Class I similarity shock solutions
to match with various asymptotic flows, Shen \& Lou (2004) modelled
(light solid lines in Fig.~\ref{fig Shen and Lou 1}) the dynamical
evolution of young stellar objects such as the Bok globule B335
system to account better for the observationally inferred density
and velocity profiles as well as the central mass accretion rate.
The downstream is part of CP 1 solution for an envelope expansion
with core collapse (EECC; Shen \& Lou 2004); this EECC solution
crosses twice the sonic critical line (viz., $x-v=1$ and
$\alpha=2/x$) at $x_{\ast}=0.23$ and $x_{\ast}=1.65$ analytically.
By choosing different shock locations $x_s=0.43,\ 1.03,\ 1.37,\ 1.43$
as in Fig.~\ref{fig Shen and Lou 1}, we readily construct various
upstream solutions to match with the analytical asymptotic solutions
at $x\rightarrow+\infty$ (see Table 1 of Shen \& Lou 2004). Comparing
with Shu et al. (2002), Shen \& Lou (2004) significantly broadened
the Class II solutions (Fig.~\ref{fig Shen and Lou 2}) for possible
`champagne flows' of H{\sevenrm II} regions around luminous OB stars
by allowing for flows at large $x$.
The upstream solutions in Shen \& Lou (2004) are not limited to the
static SIS envelope or `champagne breeze' ($V=0$); by adjusting shock
locations (or equivalently, shock speed), it is possible to match
with either asymptotic outflow (expansion or wind) solutions ($V>0$)
or asymptotic inflow (contraction or accretion) solutions ($V<0$).

\subsection{Similarity Shocks into a Static SIS Envelope}
\begin{table*}
 \centering
 \begin{minipage}{115mm}
  \caption{Class I and II similarity shock solutions
           matched with a static SIS envelope }
  \label{table solution for static}
  \begin{tabular}{@{}llllllll@{}}
  \hline
   \hline
     description&$m_0$&$x_s$&$v_d$&$\alpha_d$& $v_u$&$\alpha_u$&node\\
     \hline
$x_{\ast}(1)=0.0544$&0.105&1.26&0.466&2.00&0&1.26&1\cr
$x_{\ast}(1)=4.75\times10^{-4}$&$9.42\times10^{-4}$
                               &0.64&-0.270&3.78&0.459&3.13&2\cr
$x_{\ast}(1)=5.00\times10^{-5}$&$9.36\times10^{-5}$
                               &1.04&0.122&2.00&0&1.85&3\cr
$x_{\ast}(1)=6.50\times10^{-8}$&$1.30\times10^{-7}$
                               &0.98&-0.0227&2.04&-0.0205&2.03&4\cr
$B=7.90$&-&1.34&0.586&2.00&0&1.12&0\cr
$B=930$&-&0.63&-0.275&3.85&-0.471&3.17&1\cr
$B=9.50\times10^{4}$&-&1.04& 0.0841&2.00&0&1.85&2\cr
$B=1.10\times10^{7}$&-&0.97&-0.0338&2.07&-0.0349&2.06&3\cr

\hline
\end{tabular}
\medskip

Columns 1 through 8 provide properties of solutions ($x_{\ast}$
for Class I solutions and $B$ for Class II solutions); reduced
central mass accretion rate $m_0$ for Class I solutions; the
shock location $x_s$; the downstream velocity $v_d$; the
downstream density $\alpha_d$; the upstream velocity $v_u$, the
upstream density $\alpha_u$ and the number of stagnation points.
\end{minipage}
\end{table*}

\begin{figure}
\begin{center}
\includegraphics[scale=0.40]{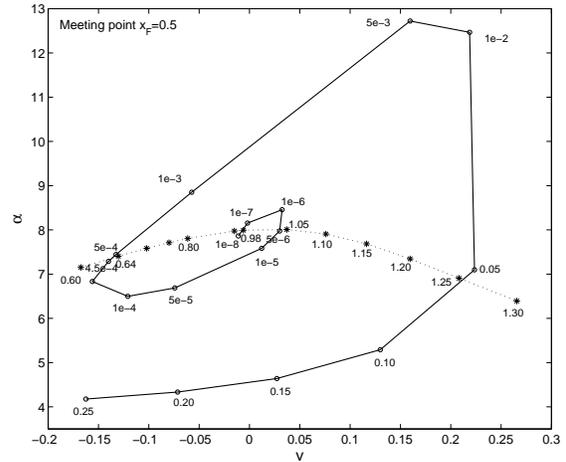}
\caption{The phase diagram of $\alpha$ versus $v$ at the meeting
point $x_{\rm F}=0.5$. Each open circle symbol indicates an
integration from $x_{\ast}$ with its value shown explicitly and
each asterisk symbol indicates an integration from the shock
location $x_{s}$ with its value also shown explicitly. The solid
and dotted curves intersect; here we only show four intersections,
from right to left, corresponding to four pairs of \{$x_{\ast}(1)$,
$x_s$\} as \{0.0554, 1.26\}, \{$5.00\times10^{-5}, 1.04$\},
\{$6.50\times10^{-8}$, 0.980\} and \{$4.75\times10^{-4}$, 0.640\},
respectively. The trend of each curve suggests that there may
exist an infinite number of discrete Class I shock solutions
matched with a static SIS envelope with $A=2$. }
\label{fig class 1 match}
\end{center}
\end{figure}

\begin{figure}
\begin{center}
\includegraphics[scale=0.45]{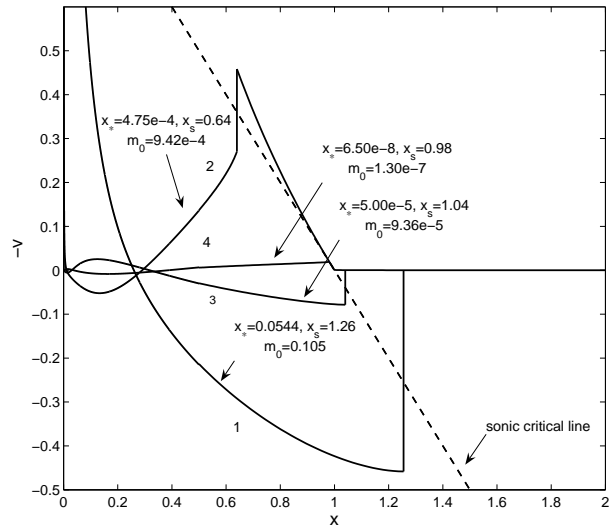}
\caption{Negative reduced radial speed $-v(x)$ versus
$x$ for the four shock solutions identified in the $\alpha-v$
phase diagram of Fig.~\ref{fig class 1 match}. The straight
dashed line is the sonic critical line. Numerals 1, 3, 4, 2
marked along the four solid curves indicate the number of
stagnation points corresponding to the shock solutions
obtained in Fig.~\ref{fig class 1 match} from right to left.
Key parameters are shown in the plot. All these shock solutions
diverge, approaching $-\infty$ as $x\rightarrow0^{+}$; such
diverging behaviours are displayed more explicitly in
Fig.~\ref{fig class 1 enlarge} below with the $x-$axis in
the logarithmic scale. }
\label{fig Class 1}
\end{center}
\end{figure}

\begin{figure}
\begin{center}
\includegraphics[scale=0.45]{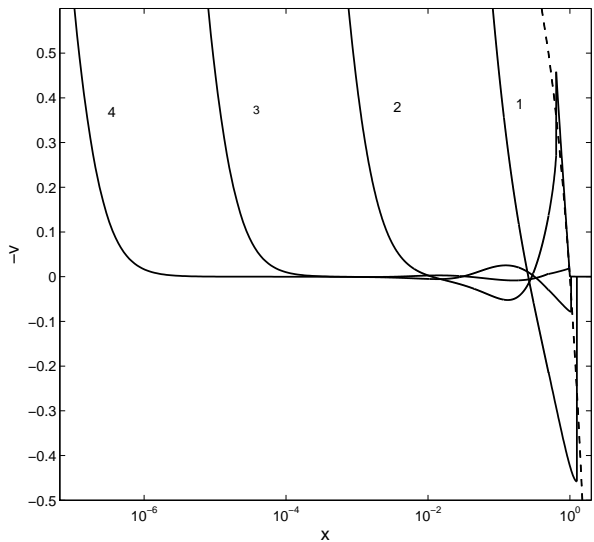}
\caption{Enlarged portions of Class I shock solution curves near
$x\rightarrow 0^{+}$ of Fig.~\ref{fig Class 1}, emphasizing the
diverging and oscillatory behaviours of this family of shock
solutions as $x\rightarrow0^{+}$. The $x-$axis is shown in the
logarithmic scale. The dashed line on the right is the sonic
critical line. The undulatory profiles of the curves marked
by numerals 2, 3, 4 represent self-similar subsonic radial
oscillations with two, three, four stagnation points, respectively. }
\label{fig class 1 enlarge}
\end{center}
\end{figure}

\begin{figure}
\begin{center}
\includegraphics[scale=0.40]{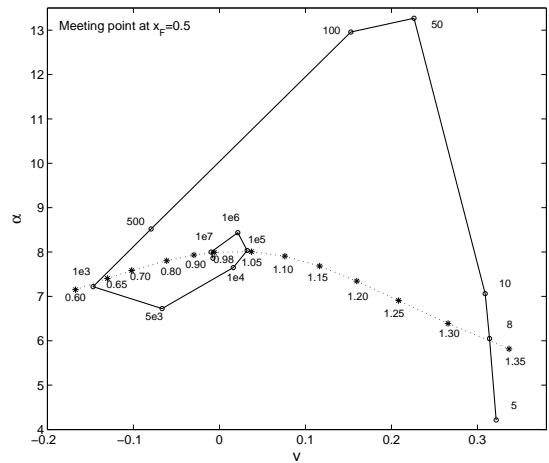}
\caption{The phase diagram of $\alpha$ versus $v$ at a chosen
meeting point $x_{\rm F}=0.5$. Each open circle symbol indicates
an integration from $x\rightarrow 0^{+}$ [e.g., $x=10^{-5}$ to
start solution (\ref{eq LP solution})] with the central reduced
mass density $B$ shown explicitly and each asterisk symbol
indicates an integration from the shock location $x_{s}$ with its
value marked explicitly. The solid and dotted curves intersect;
the first four intersections, from right to left, correspond to
the parameter pair \{$B$, $x_s$\} as \{7.90, 1.335\},
\{$9.50\times10^4$, 1.04\}, \{$1.10\times10^7$, 0.966\} and \{930,
0.633\}, respectively. The variation trend of the phase curves
suggest an infinite number of discrete Class II shock solutions
matched with a static SIS envelope $A=2$.}
\label{fig class 2 match}
\end{center}
\end{figure}

\begin{figure}
\begin{center}
\includegraphics[scale=0.45]{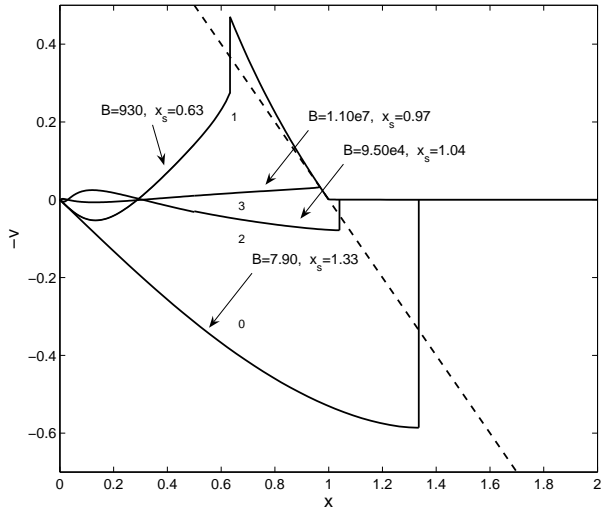}
\caption{Negative reduced speed $-v(x)$ versus
$x$ of the four shock solutions for the four intersections
in the $\alpha-v$ phase diagram in Fig.~\ref{fig class 2 match}.
The straight dashed line is the sonic line. Numerals 0, 2, 3, 1
along the four solid curves denote the number of stagnation points in
the four shock solutions determined in Fig.~\ref{fig class 2 match}
from right to left. Downstream portions are LP-type solutions. Key
parameters are shown in the plot. }
\label{fig class 2}
\end{center}
\end{figure}

\begin{figure}
\begin{center}
\includegraphics[scale=0.43]{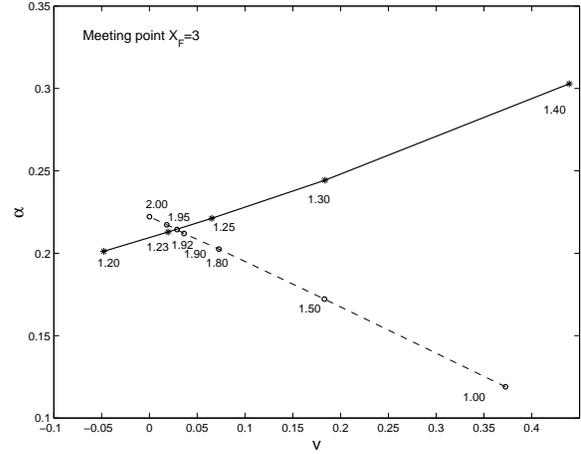}
\caption{The phase diagram of $v$ and $\alpha$ at a chosen meeting
point $x_{\rm F}=3$ for the shock solution with $x_{\ast}=0.5$. Each
asterisk symbol indicates an integration from $x_s$ with its value
shown explicitly and each open circle denotes an integration from
$x\rightarrow+\infty$ (actually from $x=20$) with the mass density
parameter $A$ also marked explicitly. From the intersection of the
solid and dashed curves, we immediately determine the shock location
$x_s$ and the mass density parameter $A$ for this shock solution. }
\label{fig class 1 breeze match}
\end{center}
\end{figure}

\begin{figure}
\includegraphics[scale=0.43]{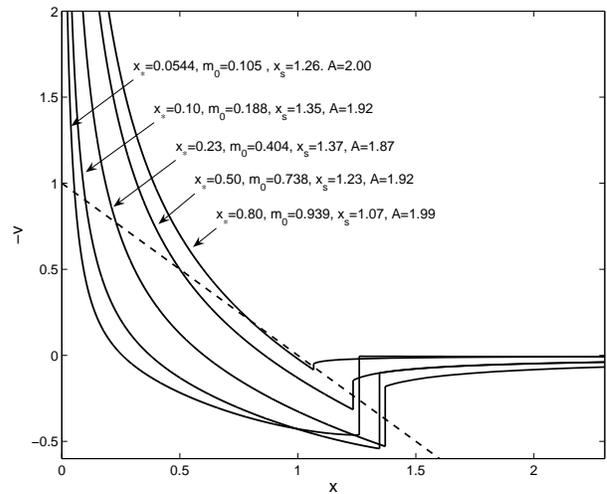}
\caption{The Class I similarity shock solutions with $V=0$. With
one stagnation point, there is only one solution matched with a
static SIS envelope. For different $x_{\ast}$, from left to right,
the solid curves represent solutions of $x_{\ast}=0.0544,\ 0.10,\
0.23,\ 0.50,\ 0.80$. Note that the shock locations $x_s$ do not
vary regularly by shifting $x_{\ast}$. Details of this class of
solutions are summarized in Table \ref{table solution for breeze}.}
\label{fig class 1 breeze}
\end{figure}

In Tsai \& Hsu (1995) and Shu et al. (2002), two classes of
solutions were derived to match with the static SIS envelope
with $V=0$ and $A=2$ or $A<2$. Here, we further explore Class
I and Class II solutions matched with the static SIS envelope
by searching possible solutions in the phase diagram of $v$ and
$\alpha$. This procedure, routinely tested earlier, was first
introduced by Hunter (1977; see Lou \& Shen 2004 for the EECC
solutions). The main technical difference is that in the prior
analyses the adjustable parameters are points along the sonic
critical line [i.e., horizontal coordinates $x_{\ast}(1)$ and
$x_{\ast}(2)$ as in Lou \& Shen (2004)], while in our cases,
more parameters are involved including $x_{\ast}(1)$, the shock
location $x_s$, the mass parameter $A$ and the velocity parameter
$V=0$. We take one parameter as fixed and adjust the other two
to match solutions in the phase diagram of $v$ and $\alpha$.

We first focus on Class I shock solutions with a static SIS envelope,
downstream solutions diverging at the centre, and a mass parameter
$A=2$ for upstream solutions. Tsai \& Hsu (1995) derived a solution
in this case and we will explore more thoroughly the phase diagram
of $v$ and $\alpha$ to construct more similarity shock solutions. In
Fig.~\ref{fig class 1 match}, we display the relevant phase diagram
of $v$ and $\alpha$ where we choose a meeting point at $x_{\rm F}=0.5$
and integrate from $x_{\ast}$(1) forward with a type 2 eigensolution
solution to $x_{\rm F}=0.5$. Meanwhile, we use the shock jump condition
to determine the $v_d$ and $\alpha_d$ as the starting condition to
integrate backward from a chosen $x_s$ to $x_{\rm F}=0.5$. For $x>x_s$,
we simply set $v=0$ and $A=2$ for a static SIS envelope. Based on Fig.
\ref{fig class 1 match}, in addition to the solution of Tsai \& Hsu,
three more similarity shock solutions belonging to this class can be
found and are shown for $-v$ versus $x$ in Fig.~\ref{fig Class 1}
(the enlarged versions for small $x$ are shown in logarithmic scale in
Fig.~\ref{fig class 1 enlarge}). The variation trend of spiral path in
the phase diagram suggests that as a class, there exists an infinite
number of discrete similarity shock solutions that connect inner
free-fall collapses to an outer static SIS envelope. Each of such
solutions is expected to be uniquely associated with a number of
stagnation points. From the intersections of the dotted and solid
curves in Fig.~\ref{fig class 1 match}, we obtain four shock
solutions, from right to left, corresponding to the displayed
solutions marked by 1, 3, 4, 2 in Figs.~\ref{fig Class 1} and
\ref{fig class 1 enlarge}; these numerals denote the number of
stagnation points, and the relevant parameters \{$x_{\ast}(1)$,
$x_s$\} are \{0.0554, 1.26\}, \{$5.00\times10^{-5}, 1.04$\},
\{$6.50\times10^{-8}$, 0.980\} and \{$4.75\times10^{-4}$, 0.640\},
respectively. The solution containing only one stagnation point is
just the solution first found by Tsai \& Hsu (1995). We find three
new solutions by exploring the phase diagram of $v$ and $\alpha$. In
principle, we can construct more solutions following the sequence
with increasing number of stagnation points. Behaviours of these
solutions at $x\rightarrow 0^{+}$ are displayed in
Fig.~\ref{fig class 1 enlarge} with velocity profiles crossing the
line $v=0$ at different $x$ values. As the value of $x_{\ast}$
becomes smaller, the number of stagnation points increases
accordingly. For example, the shock in the shock solution with
two stagnation points is an accretion shock such that both
downstream and upstream fluids move towards the centre.

We now turn to the Class II shock solutions that connect downstream
LP solutions with a static SIS envelope. Tsai \& Hsu (1995) first
obtained a solution in this case; we shall demonstrate by examples
the existence of more such Class II solutions with increasing number
of stagnation points. Fig.~\ref{fig class 2 match} is a relevant
phase diagram of $v$ and $\alpha$ to identify shock solution matches.
Here, we choose again a meeting point at $x_{\rm F}=0.5$ and integrate
forward from $x\rightarrow 0^{+}$ (e.g., $x=10^{-5}$) where the
analytical asymptotic LP solution (\ref{eq LP solution}) is imposed
with different values of the central reduced density $B$. Meanwhile,
we use the shock jump condition to determine $v_d$ and $\alpha_d$ as
the starting condition to integrate from $x_s$ back to $x_{\rm F}=0.5$.
For $x>x_s$, we simply set $v=0$ and $A=2$. The variation trend of
spiral structure in Fig.~\ref{fig class 2 match} is qualitatively
similar to that in Fig.~\ref{fig class 1 match}, pointing to a class
of infinitely many discrete similarity shock solutions that connect
inner LP solutions to an outer static SIS envelope. Based on the
phase diagram of Fig.~\ref{fig class 2}, the four intersections of
the dotted and solid curves give rise to four Class II shock
solutions, from right to left, corresponding to solutions marked by
the numerals $0, 2, 3, 1$ in Fig.~\ref{fig class 2}; these numerals
represent uniquely the number of stagnation points to identify
solutions with corresponding parameters \{$B$, $x_s$\} given by
\{7.90, 1.34\}, \{$9.50\times10^4$, 1.04\}, \{$1.10\times10^7$,
0.966\} and \{930, 0.633\}, respectively; the solution with no
stagnation point is just the solution found by Tsai \& Hsu (1995).
As $x\rightarrow 0^{+}$, the self-similar oscillation behaviours
of downstream solutions of these Class II shock solutions are
similar to those of Class I shock solutions, and as the central
reduced density $B$ becomes larger, the number of stagnation points
increases. The main difference is that as $x\rightarrow 0^{+}$,
Class II shock solutions remain finite while Class I shock
solutions diverge for core collapses.

We now summarize the main results of this section. By numerical
exploration and analysis, we derived discrete shock solutions
matched with a static SIS envelope of $V=0$ and $A=2$ within the
semi-complete space of $0<x<+\infty$ (i.e., for $x>x_s$, we set $v=0$
and $A=2$). Given this SIS envelope as the upstream condition, we
explore classes of infinite number of discrete solutions for both Class
I and Class II solutions whose downstreams are free-fall collapses and
LP solutions, respectively. For Class I solutions as $x\rightarrow0^{+}$,
the radial velocity and density profiles have asymptotic diverging
behaviours (\ref{eq free-fall}), while for Class II solutions as
$x\rightarrow 0$, the radial velocity and density profiles have
finite asymptotic behaviours (\ref{eq LP solution}) characterized
by a central reduced density $B$.
For comparison, the behaviours $x\rightarrow 0^{+}$ of this family
of shock solutions with an outer static SIS envelope are similar to
those of EECC solutions, viz., both have subsonic oscillations in
a self-similar manner as $x_{\ast}\rightarrow 0^{+}$. For Class I
solutions as the value of $x_{\ast}$ decreases, the number of
stagnation points where $v=0$ along the $x-$axis increases, while
for Class II solutions as the value of reduced density $B$ increases,
the number of stagnation points increases. The Class I and Class II
shock solutions shown by Tsai \& Hsu (1995) which contain only one
and no stagnation point along $x-$axis, respectively, are special
examples belonging to these two classes of shock solutions.

\subsection{Similarity Shock Solutions of Breeze }
\begin{table}
 \centering
 \begin{minipage}{85mm}
  \caption{Class I shock solutions with an asymptotic breeze}
  \label{table solution for breeze}
  \begin{tabular}{@{}llllllll@{}}
  \hline
   \hline
     $x_{\ast}$&$m_0$&$A$&$x_s$&$v_d$&$\alpha_d$& $v_u$&$\alpha_u$\\
\hline 0.0554&0.105&2.00&1.26&0.463&2.00&0&1.26\cr
0.10&0.188&1.92&1.35&0.572&1.72&0.102&1.11\cr
0.23&0.404&1.87&1.37&0.529&1.54&0.181&1.09\cr
0.50&0.738&1.92&1.23&0.315&1.63&0.145&1.38\cr
0.80&0.939&1.99&1.07&0.0841&1.87&0.0496&1.81\cr \hline
\end{tabular}
\medskip

Columns 1 through 8 list the values of $x_{\ast}(1)$ where
the downstream solution crosses the sonic critical line, the
reduced central mass accretion rate $m_0$, the mass density
parameter $A$ for the upstream solution, the shock location
$x_s$, the downstream velocity $v_d$, the downstream density
$\alpha_d$, the upstream velocity $v_u$ and the upstream
density $\alpha_u$, respectively.
\end{minipage}
\end{table}

In this section, we allow mass parameter $A$ to vary (i.e.,
$A\neq 2$) for a fixed $x_{\ast}$ corresponding to a specific
reduced central mass accretion rate $m_0$. Parameters $A$
and $x_s$ are adjusted to match the upstream and downstream
solutions in the phase diagram of $v$ and $\alpha$. We
integrate from a fixed $x_{\ast}$ (e.g., $0.5$ as shown in
Fig.~\ref{fig class 1 breeze match}) with a type 2 eigensolution
to the shock location $x_s$ and get the $v_d$ and $\alpha_d$.
Across the shock, we get $v_u$ and $\alpha_u$ as the starting
condition to integrate forward to a chosen meeting point
$x_{\rm F}=3$; meanwhile, we integrate back from upstream
$x\rightarrow+\infty$ by varying the mass parameter $A$ with
a fixed speed parameter $V=0$.

Here, we apply the analytical asymptotic solution (i) as given
by equation (\ref{eq asymptotic solution x-infty}) at $x=20$ as
the starting condition to integrate back to $x_{\rm F}$=3. In
Fig.~\ref{fig class 1 breeze match}, we take $x_{\ast}=0.5$ as
an example to illustrate the procedure of determining the shock
location and the mass parameter $A$ for the upstream solution in
the phase diagram of $v$ and $\alpha$. By doing so, we derive a
new family of Class I shock solutions. For example, we pick values
of $x_{\ast}$ as 0.0544, 0.1, 0.23, 0.5 and 0.8, corresponding to
values of $m_0=0.105,$ 0.188, 0.404, 0.738, and 0.939, respectively.
We find that for Class I shock solutions the value of shock velocity
and strength and the mass parameter $A$ do not vary with $x_{\ast}$
regularly as do those for Class II shock solutions. Class I shock
solutions with different $x_{\ast}$ are shown in terms of $-v$
and $x$ in Fig.~\ref{fig class 1 breeze}; in Table
\ref{table solution for breeze}, we summarize these solution results.
For example, the downstream solutions with $x_{\ast}$=0.1 and 0.5 can
match with the upstream solutions for the same mass parameter $A$.

To sum up, we present in this section the similarity solutions
with asymptotic breezes, for which, we derived a series of Class I
solutions with different central mass accretion rate $m_0$
corresponding to different values of $x_{\ast}$. This class of
solutions can be applied to processes of cloud collapse in star
formation. One important feature of these solutions is that the
value of $m_0$ is adjustable within the range of 0 to 0.975
corresponding to different central mass accretion rates or shock
locations. In one aspect, this family of solutions is analogous
to the Class II shock solutions obtained by Shu et al. (2002),
i.e., both the upstream solutions are breeze solutions. Two major
differences are: (1) our solutions are Class I shock solutions
with central free-fall collapses (i.e., diverging solutions as
$x\rightarrow 0^{+}$), while the downstream solutions of Shu et al.
(2002) are LP solutions. (2) The shock location $x_s$ and the mass
parameter $A$ for upstream solutions vary regularly as the value
of reduced central density $B$ decreases. In contrast, there is
no obvious trend of regularity for our Class I shock solutions
as $x_{\ast}$ decreases.

\subsection{Self-Similar Twin Shock Solutions}
\begin{table}
\begin{center}
 \begin{minipage}{85mm}
  \caption{Class I and II twin shock
solutions for different $x_{\ast}(2)$ }
  \begin{tabular}{@{}llllllll@{}}
  \hline
   \hline
     $x_{\ast}(2)$&Description&$x_s(1)$&$x_s(2)$&$A$&$V$\\
\hline
0.80&$x_{\ast}(1)=3.40\times10^{-4}$&0.670&0.85&1.43&-0.528\cr
&$m_0=6.20\times10^{-4}$&&&\cr &$B=1300$&0.660&0.90&1.52&-0.430\cr
&&&1.00&1.77&0.192\cr &&&1.07&1.99&0\cr &&&1.10&2.12&0.109\cr
&&&1.15&2.36&0.294\cr
0.90&$x_{\ast}(1)=4.40\times10^{-4}$&0.640&0.95&1.77&-0.198\cr
&$m_0=8.18\times10^{-4}$&&&\cr &B=1000&0.627&1.00&1.91&-0.0708\cr
&&&1.05&2.10&0.0831\cr \hline
\end{tabular}
\medskip

Columns 1 through 6 list values of $x_{\ast}(2)$ where the Class I
twin shock solutions cross the sonic critical line the second time,
the key parameters [$x_{\ast}(1)$ and $m_0$ for Class I and $B$
for Class II solutions, respectively], the first shock location
$x_s(1)$, the second shock location $x_s(2)$, the mass density
parameter $A$ for the upstream solution and the speed parameter
$V$, respectively.
\end{minipage}
\end{center}
\end{table}

\begin{figure}
\begin{center}
\includegraphics[scale=0.45]{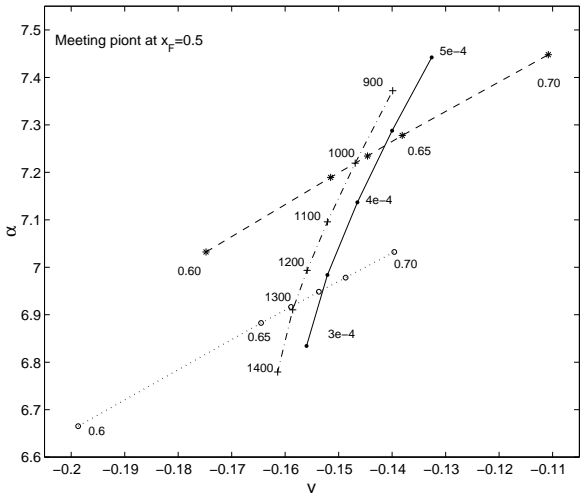}
\caption{The phase diagram of $\alpha$ versus $v$ at a chosen meeting
point $x_{\rm F}=0.5$. Each solid dot denotes an integration from
$x_{\ast}(1)$ (its value shown explicitly) with type 2 eigensolution,
each cross symbol denotes an integration from $x\rightarrow 0^{+}$ (e.g.,
$x=10^{-5}$) with the value of the central reduced mass density B shown
explicitly, and each open circle and asterisk symbols denote integrations
from $x_{\ast}(2)=0.8$ and $x_{\ast}(2)=0.9$ respectively via shock
condition to the meeting point $x_{\rm F}=0.5$; numerals indicate the
shock location $x_s$. From the curve intersections, we determine the
Class I and II shock solutions for $x_{\ast}=0.8$ and $0.9$, respectively. }
\label{fig 0809 match}
\end{center}
\end{figure}

\begin{figure}
\begin{center}
\includegraphics[scale=0.43]{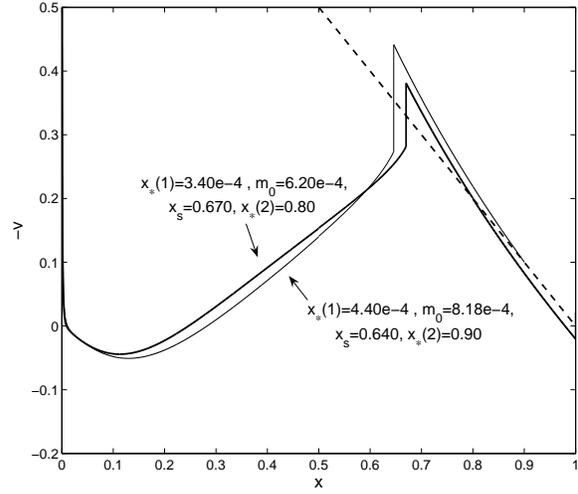}
\caption{Class I shock solutions with $x_{\ast}(2)=0.8$ and $0.9$
for $-v(x)$ versus $x$ denoted by the light
and heavy solid curves, respectively.
The straight dashed line is the sonic critical line.
The two corresponding values of $x_{\ast}(1)$ are
$4.40\times 10^{-4}$ and $3.40\times 10^{-4}$.  }
\label{fig class 1 0809}
\end{center}
\end{figure}

\begin{figure}
\begin{center}
\includegraphics[scale=0.43]{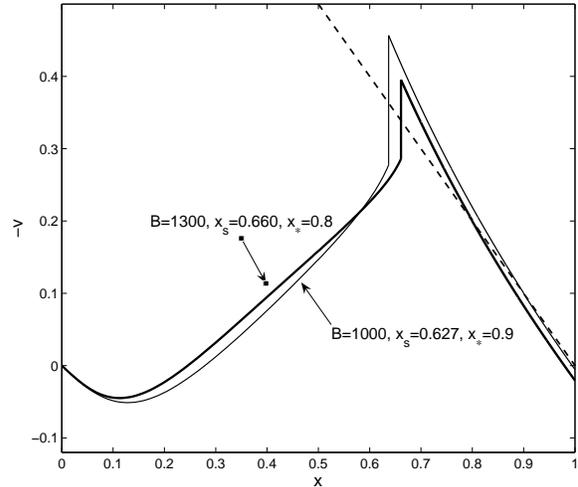}
\caption{Class II shock solutions with $x_{\ast}=0.8$ and $0.9$
for $-v(x)$ versus $x$ denoted by the light and heavy solid
curves, respectively. The straight dashed line is the sonic
critical line. Key parameters are shown in the plot. }
\label{fig class 2 0809}
\end{center}
\end{figure}

\begin{figure}
\begin{center}
\includegraphics[scale=0.43]{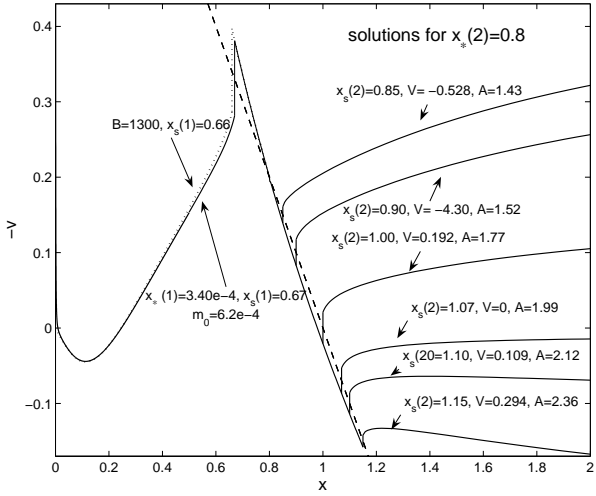}
\caption{Class I and II twin shock solutions with $x_{\ast}=0.8$
represented by the solid and dotted curves, respectively. The
straight dashed line is the sonic critical line. Once the upstream
solution of the first shock at $x_s(1)$ crosses the sonic critical
line analytically at $x_{\ast}=0.8$ with a type 2 eigensolution,
we can choose the second shock location $x_s(2)=0.85$, 0.90, 1.00,
1.07, 1.10, 1.15 to match with various asymptotic solutions at
$x\rightarrow +\infty$.  } \label{fig 08}
\end{center}
\end{figure}

\begin{figure}
\begin{center}
\includegraphics[scale=0.43]{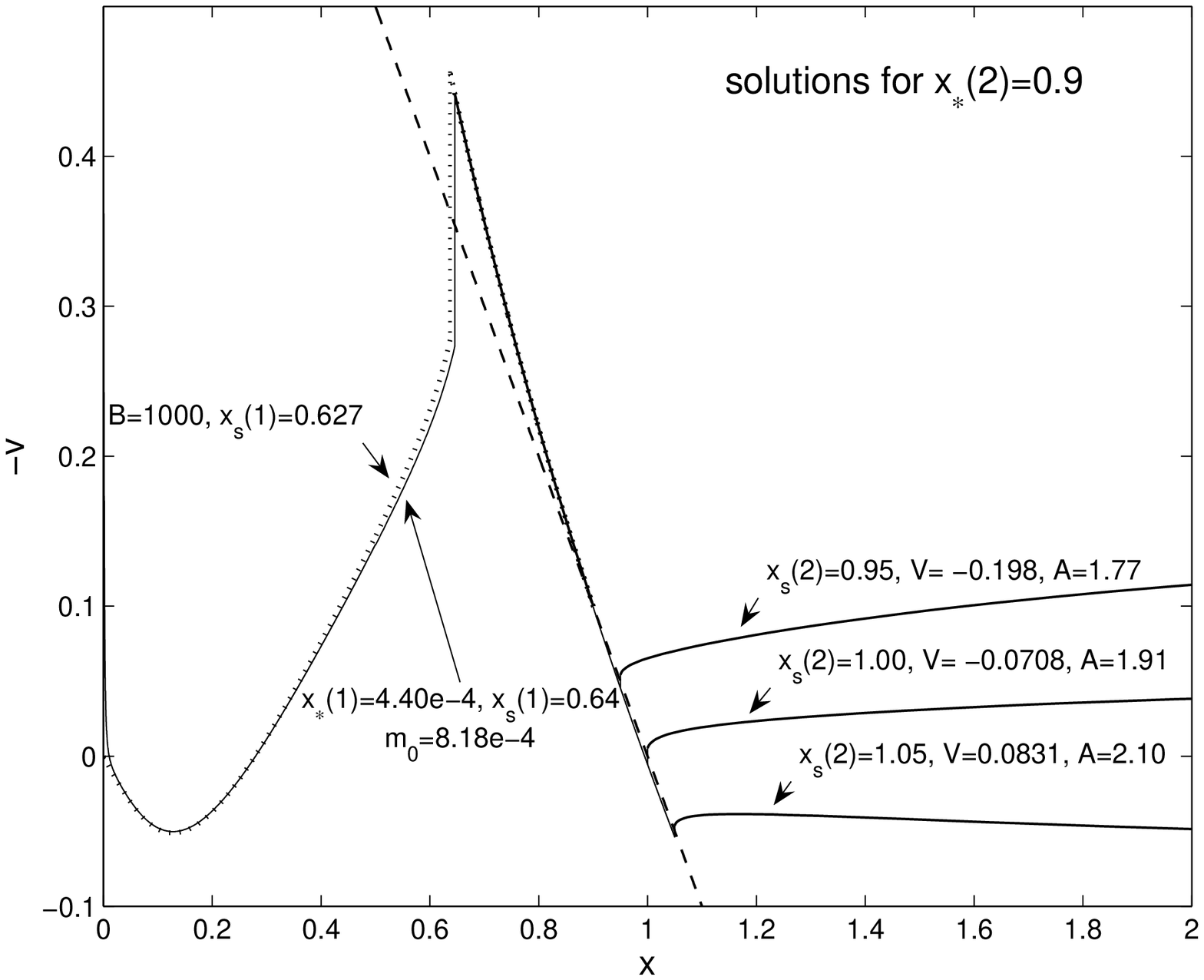}
\caption{Class I and II twin shock solutions with $x_{\ast}=0.9$
represented by the solid and dotted curves, respectively. The straight
dashed line is the sonic critical line. Once the upstream solution of
the first shock at $x_s(1)$ crosses the sonic critical line analytically,
we choose for example $x_s(2)=0.95, 1.00, 1.05$ for the second shock
location to match with asymptotic solution at $x\rightarrow +\infty$. }
\label{fig 09}
\end{center}
\end{figure}

In all previous studies, the similarity shock solutions contain only
one shock and the Class I solutions crosses the sonic critical line
only once analytically. In this section, we construct similarity
shock solutions with twin shocks.

In \S\ 3.2, we find a Class I shock solution crossing the sonic critical
line at $x_{\ast}=4.75\times10^{-4}$ with a shock location $x_s=0.640$
and two stagnation points (see Fig.~\ref{fig Class 1}), and a Class II
shock solution with a central reduced density $B=930$ and a shock
location at $x_s=0.633$ (Fig.~\ref{fig class 2}) containing only one
stagnation point (n.b., the one of $v\rightarrow 0$ as
$x\rightarrow 0^{+}$ is not counted here). The portions of upstream
solutions between the shock and the static SIS envelope of both shock
solutions are tangent to the sonic critical line at the point of
$x_{\ast}(2)=1$. In fact, the upstream solution is part of the EWCS
(Shu 1977). As we shift the $x_{\ast}(2)$ location leftward into the
interval $0<x<1$, the upstream solution of the first (i.e., leftmost)
shock may cross the sonic critical line analytically with a type 2
eigensolution. We can then construct Class I and Class II twin shock
solutions in the $\alpha -v$ phase diagram given this real possibility.

Let us first focus on Class I twin shock solutions. We fix the
value of $x_{\ast}(2)$ and adjust the value of $x_{\ast}(1)$ and
the shock location $x_s(1)$ between $x_{\ast}(1)$ and
$x_{\ast}(2)$ to find the similarity shock solution crossing the
sonic critical line twice analytically based on intersections in
the $\alpha-v$ phase diagram. On this ground, it is possible to
further construct a Class I similarity solution with twin shocks
and an appropriate asymptotic solution at $x\rightarrow + \infty$. We
take a meeting point at $x_{\rm F}=0.5$ and integrate an upstream
solution of the first shock from $x_{\ast}(2)$ for Class I solutions
by using a type 2 eigensolution as a starting condition to a chosen
shock location $x_s(1)$. Based on the velocity and density upstream
of this point \{$v_u$, $\alpha_u$\}, we use the shock jump conditions
(\ref{eq shock condition (1) for tau=1}) and
(\ref{eq shock condition (2) for tau=1}) to determine the velocity
and density \{$v_d$, $\alpha_d$\} downstream of $x_s(1)$ and to
integrate from $x_s(1)$ to $x_{\rm F}=0.5$; meanwhile, we integrate
forward from $x_{\ast}(1)$ using another type 2 eigensolution to
$x_{\rm F}=0.5$. In Fig.~\ref{fig 0809 match}, we show the phase
diagram for the two cases of $x_{\ast}(2)=0.8$ and $0.9$, respectively.
From the intersections of the solid curve and the dotted curve as well
as the dashed curve, we obtain the relevant parameter pair
\{$x_{\ast}(1)$ $x_s(1)$\} as \{$3.40\times10^{-4}$, 0.670\} for the
solution with $x_{\ast}(2)=0.8$ and as \{$4.40\times10^{-4}$, 0.640\}
for the solution with $x_{\ast}(2)=0.9$. We show the corresponding
solutions for $-v$ versus $x$ in Fig.~\ref{fig class 1 0809}.

For Class II solutions, we match the upstream and downstream solutions
in the $\alpha-v$ phase diagram using a search procedure similar to
that of finding the Class I twin shock solutions. Given a fixed
$x_{\ast}$, we adjust the value of central reduced mass density $B$
and the shock location $x_s(1)$ in order to find matches. At a chosen
meeting point $x_{\rm F}=0.5$, we first obtain different parameter sets
\{$v$, $\alpha$\} denoted by cross symbols for different $B$ values by
integrating forward from $x\rightarrow 0^{+}$ with the LP solution. We
then integrate backward from $x_{\ast}$ to an adjustable shock location
$x_s(1)$; by imposing the shock condition, we continue to integrate
backward towards $x_{\rm F}=0.5$ to obtain a pair of \{$v$, $\alpha$\}.
In Fig.~\ref{fig 0809 match}, we thus obtain open circle symbols for
the case of $x_{\ast}=0.8$ with different $x_s(1)$ and asterisk symbols
for the case of $x_{\ast}=0.9$ with different $x_s(1)$. From the
intersections of the dash-dotted curve with the dotted and dashed
curves, we find the parameter pair \{$B$, $x_s(1)$\} as \{1300, 0.660\}
and \{1000, 0.627\} for the shock solutions of $x_{\ast}=0.8$ and
$x_{\ast}=0.9$, respectively.

By matching the downstream and upstream solutions in the phase
diagram, we obtain the solution structure around the first shock
$x_s(1)$. For a solution passing the sonic critical line at
$x_{\ast}$ the second time, we can further construct the second
shock by choosing the different location $x_s(2)$ of the second
shock. For $x_{\ast}=0.8$, we choose $x_s(2)=0.85$, 0.90, 1.00,
1.07, 1.10 and 1.15 while for $x_{\ast}=0.9$, we choose
$x_s(2)=0.95$, 1.00, 1.05. The upstream solutions across this
second shock can match with the analytical asymptotic solution
(\ref{eq asymptotic solution x-infty}) at $x\rightarrow +\infty$;
the property of these upstream solutions can evolve from the
inflow (contraction or accretion) to outflow (expansion or wind
or breeze), as the value of the shock location $x_s(2)$ increases.

The main properties of the similarity twin shock solutions are: (1)
there exist two stagnation points for the Class I twin shock solutions,
while there is only one stagnation point for the Class II twin shock
solutions excluding $x\rightarrow 0^{+}$; the spherical stagnation
surfaces of zero flow speed travel with different yet constant
subsonic speeds, indicating subsonic radial oscillations; (2) the first
shock near the core is an accretion shock, as the reduced radial speeds
of the pre-shock and post-shock are both negative for inflows. This
accretion shock expands at a constant subsonic speed. For the same
$x_{\ast}(2)$ and $x_{\ast}$, the velocity of the first shock in Class I
solution is a bit larger than that in Class II solution, and for the same
class of shock solutions, the velocity of the first shock becomes larger
as the $x_{\ast}(2)$ becomes smaller; and (3) the location of the second
shock $x_{s}(2)$ can be larger or smaller than 1, such that the velocity
of this shock can be supersonic or subsonic. The upstream solution for
this shock evolves from the accretion condition to breeze to wind. For
different astronomical system, we can construct different shock solutions
(e.g., an envelope contraction with core collapse or an envelope expansion
with core collapse).

\subsection{Two-Temperature Similarity Shock Solutions }

\begin{table*}
 \centering
 \begin{minipage}{100mm}
  \caption{Class I and II Shock Breeze Solutions
                    with Different $\tau$ Values}
  \label{table for breeze different tau}
  \begin{tabular}{@{}lllllllll@{}}
  \hline
   \hline
     Parameters &$\tau$&$A$&$x_{sd}$&$x_{su}$&$v_d$
                &$\alpha_d$& $v_u$&$\alpha_u$\\
\hline
$x_{\ast}(1)=0.23$&1&1.86&1.37&1.37& 0.529&1.54&0.181&1.09\cr
&1.3&2.66&1.34&1.74& 0.515&1.58&-0.450&0.770\cr
&1.5&3.31&1.31&1.62&0.500&1.62&-0.731&0.732\cr
&1.8&4.49&1.27&2.29&0.480&1.69&-1.12&0.704\cr
$B=0.10$&1&0.23&2.42&2.42&1.76&0.173&0.898&0.0746\cr
&1.3&0.48&2.33&3.03&1.69&0.166&0.575&0.0564\cr
&1.5&0.58&2.29&3.44&1.66&0.164&0.452&0.0521\cr
&1.8&0.78&2.25&4.05&1.62&0.161&0.316&0.0485\cr \hline

\end{tabular}
\medskip

Columns 1 to 9 contain shock solution properties [$x_{\ast}(1)$
for Class I solutions and $B$ for Class II solutions]; the ratio
$\tau$ of the downstream ($a_d$) to upstream ($a_u$) sound speed;
the mass parameter $A$ for the upstream solution; the downstream
shock location $x_{sd}$; the upstream shock location $x_{su}$;
the downstream reduced velocity $v_d$; the downstream reduced
density $\alpha_d$; the upstream reduced velocity $v_u$ and the
upstream reduced density $\alpha_u$, respectively.
\end{minipage}
\end{table*}

\begin{table*}
 \centering
 \begin{minipage}{110mm}
  \caption{Class II Shock Solutions for $\tau=1.5$ with $B=10^{-6}$. }
  \label{table for tau=1.5 for diff x_s}
  \begin{tabular}{@{}lllllllll@{}}
  \hline
   \hline
    $x_{sd}$&$x_{su}$&$v_d$&$\alpha_d$&$v_u$&$\alpha_u$&$A$&$V$\\
\hline 2.0&3.0&1.44&$1.58\times10^{-6}$&-0.209&$4.12\times10^{-7}$
&$8.41\times10^{-7}$&-0.954\cr
3.0&4.5&2.30&$2.73\times10^{-6}$&1.65&$1.00\times10^{-6}$
&$4.65\times10^{-6}$&0.998\cr
3.5&5.25&2.76&$3.79\times10^{-6}$&2.47&$1.51\times10^{-6}$
&$9.61\times10^{-6}$&1.85\cr

\hline
\end{tabular}
\medskip

Columns 1 to 8 are the downstream shock location $x_{sd}$;
the upstream shock location $x_{su}$; the downstream reduced
radial speed $v_d$; the downstream reduced density $\alpha_d$;
the upstream reduced radial speed $v_u$; the upstream reduced
density $\alpha_u$ and the mass parameter $A$ and the speed
parameter $V$ for the upstream solution, respectively.
\end{minipage}
\end{table*}

\begin{figure}
\begin{center}
\includegraphics[scale=0.43]{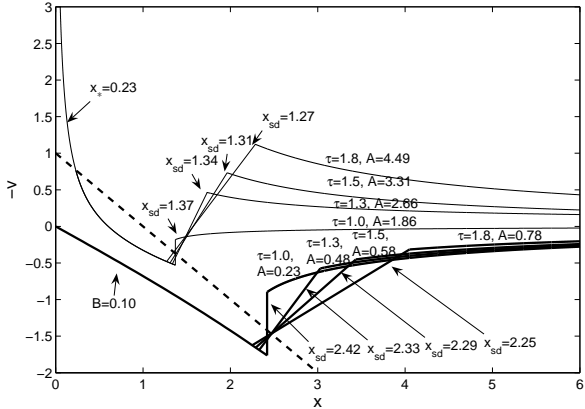}
\caption{The negative reduced radial speed $-v(x)$ versus $x$ for
$\tau\geq 1$. The straight dashed line is the sonic critical line.
The light solid curves form the Class I shock breeze or
contraction solutions with $x_{\ast}=0.23$ and different $\tau\geq
1$, and the heavy solid curves form the Class II shock breeze
solutions with $B=0.10$ for different $\tau\geq 1$. Here, we choose
$\tau$ to be 1.0, 1.3, 1.5 and 1.8 to get different shock breeze or
contraction solutions. }  \label{fig differenttau}
\end{center}
\end{figure}

\begin{figure}
\begin{center}
\includegraphics[scale=0.43]{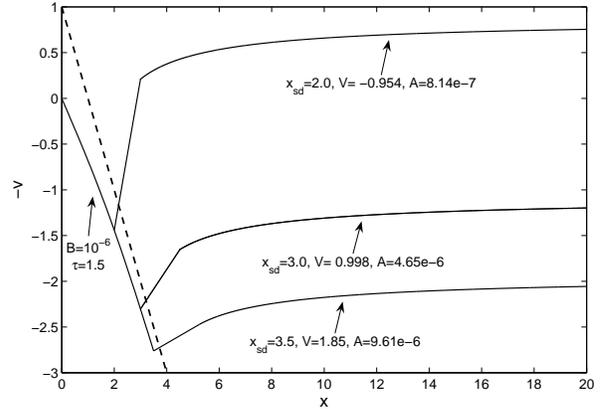}
\caption{The negative reduced radial speed $-v(x)$ versus $x$.
The straight dashed line is the sonic critical line. The solid
curves form the Class II shock solutions with $B=10^{-6}$ for
$\tau=1.5$. We choose $x_{sd}$ to be 2.0, 3.0, 3.5 to construct
different shock solutions matched with asymptotic inflow and winds. }
\label{fig differenttau lp}
\end{center}
\end{figure}

In the preceding sections, we only consider the case of $\tau=1$
corresponding to the same thermal temperature for both downstream
and upstream fluids such that the entire fluid has the same sound
speed $a_u=a_d$. For the isothermal condition, the sound speed
can be expressed as
\begin{equation}
a=\bigg(\frac{P}{\rho}\bigg)^{1/2}
=\bigg[\frac{(Z+1)k}{\mu}T\bigg]^{1/2}\ ,
\label{equa sound}
\end{equation}
where $k$ is the Boltzmann constant, and sound speed $a$ is
determined by the mean atomic mass $\mu$, the ionization state
$Z$ and the temperature $T$ together. In some astrophysical
processes, these factors determining the sound speed should
evolve from an interior region to an exterior region (e.g.,
H{\sevenrm II} regions around luminous massive OB stars); in
this sense, a fluid may not be treated as an isothermal system.
In this section, we will discuss a system of two-temperature
fluid. In such a system, the downstream sound speed $a_d$
differs from that of the upstream $a_u$. Meanwhile, the
downstream and upstream fluids can be treated as isothermal
fluids respectively. In most astrophysical systems, the
downstream sound speed should be larger than that of the
upstream. As the central objects heat up the gas around them,
we presume $\tau\equiv a_d/a_u >1$ in our numerical examples.

Let us begin with the shock solutions for the case of $V=0$ at
$x\rightarrow+\infty$. Here, we take on Class I solutions with
$x_{\ast}(1)=0.23$ and Class II solutions with $B=0.1$ as examples
of illustration. By  matching the upstream and downstream solutions
in the $\alpha-v$ phase diagram at a meeting point $x_{\rm F}$, we
obtain the Class I and Class II shock solutions for different
values of $\tau$ (e.g., $\tau=10, 1.3, 1.5, 1.8$). In
Fig.~\ref{fig differenttau}, we display both classes of shock
solutions for $-v(x)$ versus $x$. Relevant parameters are summarized
in Table \ref{table for breeze different tau}. For a prescribed
boundary condition at $x\rightarrow0^{+}$ [i.e., $m_0$ corresponding
to $x_{\ast}(1)$ or the central reduced mass density $B$], as
parameter $\tau$ increases, the downstream shock location $x_{sd}$
representing the Mach number of downstream shock becomes smaller;
meanwhile, the mass parameter $A$ of the upstream solution increases.
For $A<2$, the upstream solution gives an outflow breeze; as the value
of $A$ increases, the breeze becomes weaker. For $A>2$, an inward
contraction appears at $x\rightarrow +\infty$; the strength of this
contraction or inflow becomes larger, as the value of $A$ increases.

We then allow the velocity parameter $V\neq 0$ to vary. As an example,
we take $\tau=1.5$ to derive the Class II shock solutions. By choosing
different downstream shock locations $x_{sd}$, we construct and
display different shock solutions for $-v(x)$ versus $x$ in
Fig.~\ref{fig differenttau lp} with the reduced central density
$B=10^{-6}$. Details of such type of two-temperature shock solutions
are summarized in Table \ref{table for tau=1.5 for diff x_s}. From
this family of shock solutions, we note that with increasing value of
$x_{sd}$, the upstream solutions gradually change from an accretion
to a wind, and the shock strength becomes weaker. As $x_{sd}$ is
increased further, $v_d$ will becomes smaller than $v_u$ (e.g.,
$x_s=4$ for $\tau=1.5$) which is unphysical. That is, the downstream
shock location $x_{sd}$ cannot be arbitrarily large.

\section{Astrophysical Applications}

Self-similar solutions have been studied to model various astrophysical
systems previously (e.g., Larson 1969; Penston 1969; Shu 1977; Tsai \&
Hsu 1986; Shu et al. 1987, 2002; Lou \& Shen 2004; Shen \& Lou 2004).
There are many instances in which these solutions allow one to develop
a numerically simple description of what one first feels is a very
complicated astrophysical flow. As results of instabilities and
supersonic flows of fluid, accretion and expansion shocks are expected
to emerge in various cosmological and astrophysical settings (e.g.,
supernova explosions; dynamical  H{\sevenrm II} regions surrounding
luminous massive OB stars; the dynamic evolution from AGBs to pPNe;
accretion shocks around BHs and in quasars; possibly, the creation of
initial fireballs in a class of GRBs). In this paper, we mainly focus
on various theoretical possibilities and produce different types of
similarity shock solutions, including the solutions for a static SIS
envelope, for asymptotic breezes or winds or inflows, for twin shocks,
for two-temperature fluid and for the global feature of EECC solutions
as well as various other possible combinations. Here, we discuss and
outline potential astrophysical applications for these similarity shock
solutions.

1. The dynamic process connecting the AGB phase
and the pPN phase (e.g., Lou \& Shen 2004).
The stellar evolutionary stage between the end of the AGB phase and the
PN phase has long been a key missing link in our physical understanding
for the evolution of a single star (e.g., Balick \& Frank 2002). A
swelling and expanding giant or supergiant star with a massive yet slow
wind eventually end up with a more or less detached system containing a
small compact and extremely hot white dwarf at its center and a nebula of
various morphologies with signatures of shocks indicating interactions of
hot faster winds ($\sim 10^3\hbox{ km s}^{-1}$) with the massive slow wind
($\sim 10-20\hbox{ km s}^{-1}$). In the gross picture, a dynamic process
characterized by self-similar EECC shock solutions may be highly relevant:
expansion and outflow of gas materials surrounding a proto-white dwarf in
a pPNe, meanwhile infall and collapse in the central core region to produce
a compact and hot white dwarf.

Sufficiently far away from initial and boundary conditions, a
largely isolated stellar system may gradually evolve into a
dynamic phase in a self-similar manner. Available observations
suggest the following constraints:
(1) the timescale of this collapse-expansion stage is estimated
to be $\sim 10^3$ yrs;
(2) the mass of the central white dwarf should be less than the
Chandrasekhar limit of 1.39~M$_{\odot}$, otherwise nova-like or
even Type Ia supernova-like activities might occur sporadically
to produce anomalous abundances in PNe;
(3) the massive shell envelope expands with a typical speed of
$\sim10-20$~km~s$^{-1}$ which is comparable to the sound speed
in the gas;
(4) there are numerous indications that PNe form involving a
much slower wind from the progenitor giant star overtaken by a
subsequent faster hot wind presumably generated when a white dwarf
emerges, such that shocks are inevitable (Kwok 1978, 1982, 1993;
Balick \& Frank 2002).

Adopting highly idealized Class I similarity shock solutions
constructed in this theoretical analysis, we attempt to provide
the following rough estimates in reference to the theoretical
central mass accretion rate given by $m_0a^3/G$ where $a$ is the
isothermal sound speed and $G$ is the gravitational constant. Here,
we take the sound speed $a$ as $\sim 20$~km~s$^{-1}$ with a mass
accretion rate of the order of $\sim 2m_0$~M$_{\bigodot}$/yr. For
a timescale of $\sim 10^3$~yr, the total accumulated mass onto the
central white dwarf would be $\sim 2\times10^3m_0$~M$_{\bigodot}$.
As the mass limit of a white dwarf $1.39~M_{\bigodot}$ and
considering the possible range of sound speeds, the parameter
$m_0$ should fall in the rough range of $\sim 10^{-3}-10^{-4}$.
In our analysis, $m_0$ associate with the Class I shock solutions with
only one stagnation point appears to fit in this range. For example,
we may take a Class I twin shock solution with $x_{\ast}(1)=0.8$ with
a corresponding $m_0=6.20\times10^{-4}$; this would give a mass of
$\sim1.2$~M$_{\bigodot}$ for the eventual white dwarf. By choosing
the second shock location at $x_{s}(2)=1.15$, we can derive the
expansion speed of the second shock to be $\sim 20$~km~s$^{-1}$
and the wind speed of $\sim $6~km~s$^{-1}$, consistent with
observations of wind speed in the range of $\sim 5-10$~km~s$^{-1}$.
We note that there exists an accretion shock in the interior
region [i.e., $x_s(1)$], which expands at a constant speed of
$\sim 13$~km~s$^{-1}$; this expansive shell form part of interior
structures of a PN.

2. Cloud collapses in the star formation process. Shu et al. (1987)
reviewed star formation processes in molecular clouds. They suggested
to divide the star formation process into four stages. In this scheme,
the self-similar evolution may be applicable in relatively early
stages referred to as the pre-stellar stage of star formation defined
as the dynamic phase in which a gravitationally bound core has formed
in a molecular cloud and evolves towards higher degrees of central
condensation, but no central hydrostatic protostellar object exists
yet (e.g., Andr$\acute{\hbox{e}}$ et al. 1999). As idealization and
simplification, a molecular cloud may be treated as spherically
symmetric and isothermal during that stage.

With this, the self-similar shock solutions to be used in
star formation processes can be classified into two types:

(i) The solutions with an asymptotic radial speed $u\rightarrow 0$
at $r\rightarrow+\infty$. This type of solutions has been discussed
by Shu (1977) and Tsai \& Hsu (1995) for a static SIS envelope with
$V=0$ and $A=2$ in (\ref{eq asymptotic solution x-infty}). In \S\ 3.3,
we derive various breeze solutions with velocity and density profiles
and different values of $m_0$. For the central mass accretion rate
$\dot{M}=m_0a^3/G$, the value of $m_0$ is 0.975 for the EWCS, while
in our shock breeze solutions, the range of $0-0.975$ for $m_0$ is
much wider. This is an important theoretical leeway to account for
various central mass accretion rates in protostars. For a typical
temperature of 10~K in molecular clouds, we take a sound speed
$a=0.2$~km~s$^{-1}$ and a central mass accretion rate of
$\dot{M}\sim2\times10^{-6}m_0$~M$_{\bigodot}$/yr. As the timescale of
this stage is about $10^6-10^7$~yrs (e.g., Jessop \& Ward-Thompson 2001),
the formation of a protostar with a mass of $\sim$ 1 M$_{\bigodot}$ would
imply a parameter $m_0\gsim 0.1$.
Meanwhile as the gravitational energy is released during the core
collapse, one would expect a point source of infrared luminosity
$L_{\rm IR}\sim a^6t{m_0}^2/(GR_{\ast})$ emanating from the dense
gas cloud surrounding the core, where $R_{\ast}$ is a protostellar
radius taken to be $\sim 3$~R$_{\bigodot}$ (e.g., Stahler 1988).
Note that a factor of $m_0^2$ is involved here. For $m_0\sim 1$ as
in the case of an EWCS of Shu (1977), this $m_0^2$ factor is fairly
close to $1$. For smaller values of $m_0$, this $m_0^2$ factor may
reduce the $L_{\rm IR}$ significantly. For the dense core L1544 in
the Taurus molecular complex with a lifetime of $\sim 0.3-0.5$~Myr
(e.g., Tafalla et al. 1998) and at the distance of L1544, the
sensitivity of the {\it Infrared Astronomy Satellite (IRAS)} was
about 0.1~L$_{\bigodot}$. As the L1544 cloud system could not be
detected by the {\it IRAS}, the infrared luminosity $L_{\rm IR}$
should be less than 0.1~L$_{\bigodot}$. From this, we may infer the
upper limit of $m_0$ to be $\sim 0.1$ for the core of the L1544
cloud system. Among other factors and considerations as well as
possibilities, our shock solutions here properly adapted should
be applicable to L1544 system.
We note that Myers (2005) recently presented the
models of free-fall gravitationl collapse in the equilibrium layers,
cylinders and Bonnor-Ebert spheres, and the spherical models can
match with the observed inward velocity very well. Because of the
very little accretion rate of the collapsing Bonner-Ebert sphere
in the early stage of the star formation, the model of Bonnor-Ebert
spheres may be an alternative way of solving the luminosity puzzle
of L1544 cloud system involving centrally condensed collapse of a
starless core.

(ii) The shock solutions with a constant asymptotic radial speed
$u$ as $r\rightarrow+\infty$. Recent observations have revealed
that at sufficiently large distances away from the core, the
envelopes of some molecular clouds may expand with a speed of
$\sim 0.25$~km~s$^{-1}$, hinting at the characteristic feature
of the EECC shock solutions; these theoretical possibilities
were first discussed by Shen \& Lou (2004).

3. The ionization and expansion of H{\sevenrm II} regions. As
massive OB stars turn on their central nuclear reactions in modecular
clouds,
the interstellar medium of surrounding neutral hydrogen gas strongly
absorbs the ultraviolet radiation from OB stars and becomes ionized
to form H{\sevenrm II} regions (e.g. Str$\ddot{\hbox{o}}$mgren 1939;
Osterbrock 1989 and extensive references therein). As the ionization
front sweeps through the cloud, flows driven by pressure gradients
develop between H{\sevenrm II} and H{\sevenrm I} regions as well as
within H{\sevenrm II} regions, and shocks would conceivably emerge
in these processes. Physically similar processes but on much larger
scales are also expected to occur in distant quasars and starburst
galaxies (e.g., redshifts $z>6$) where surrounding neutral hydrogen
gas clouds are irradiated by ultraviolet photons from central accretion
or starburst activities and are ionized with subsequent dynamical
consequences. For distant quasars, features in the spectral range
between the Gunn-Peterson absorption trough (Gunn \& Peterson 1965;
Scheuer 1965) blueward of the Ly$\alpha$ emission line (e.g., Fan et
al. 2001) and the Ly$\alpha$ emission line should contain valuable
diagnostic information of the underlying dynamics of H{\sevenrm II}
regions. We shall further pursue this line of research by combining
radiative transfer processes with various shock flow models in
separate papers.
The expansion of the H{\sevenrm II} regions has been studied for
decades (e.g., Newman \& Axford 1968; Mathews \& O'Dell 1969;
Tenorio-Tagle 1979; Franco, Tenorio-Tagle \& Bodenheimer 1990;
Shu et al. 2002; Shen \& Lou 2004). The last two analyses invoked
the self-similar shock solutions to model flows and shocks in
H{\sevenrm II} regions. Because of the heating from the central object
and the ionization of the gas medium, the sound speed in the inner
and outer regions should be different in general. In our relatively
simple model framework, a two-temperature fluid separated by a shock
may grab certain gross thermal features better. For a given central
reduced mass density, by choosing a sound speed ratio of $a_{sd}$
and $a_{su}$ (e.g., $\tau=1.5$) and the shock location $x_{sd}$
($x_{su}=\tau x_{sd}$), it is possible to construct different shock
solutions to match with different asymptotic solutions (e.g.,
outflow, inflow and static envelope) at $x\rightarrow+\infty$.

We note that when the isothermal approximation is replaced by the
polytropic approximation (Suto \& Silk 1988; Lou \& Gao 2005 in
preparation; Wang \& Lou 2005 in preparation), it would be more
natural to deal with temperature variations in flows and across shocks.

4. Accretion shocks around supermassive black holes such as quasars.
Gas materials around quasars fall towards the massive centre following
the immense pull of gravity. As the infalling gas with a speed of
$\sim 500$~km~s$^{-1}$ impacts on the core gas materials, a strong
accretion shock emerges. Downstream of such a shock, neutral hydrogen
gas is expected to be fully ionized. Because of absorptions by the
infalling pre-shock gas, a sharp radiative flux drop is observed in
quasar spectra indicating the existence of accretion shocks. For the
conceptual simplicity, the geometry of such an accretion shock is
approximated as spherically symmetric. The temperature of surrounding
shock heated gas is higher than $\sim 10^{7}$~K with a sound speed of
about several $10^2$~km~s$^{-1}$. Part of the gas is expected to
collapse onto the galactic disk eventually. On larger scales, we may
apply our similarity shock solutions to model accretion shocks for
such a quasar system. In our scenario, the upstream flow velocity is
$\sim 0.5$ sound speed corresponding to several $10^2$~km~s$^{-1}$
and the shock front expands at a constant speed of several
$10^2$~km~s$^{-1}$. As the lifetime of a quasar is about $10^8$~yrs,
the radius of such a shock sphere is about 0.1~Mpc and the central
mass accretion rate can reach $10^2$ M$_{\bigodot}\hbox{ yr}^{-1}$.
This accumulation of materials can assemble the host galaxy of a
quasar in a timescale of several $10^8$ yrs. These estimates are
well consistent with the results of Barkana \& Loeb (2003). Generally
speaking, in the reservoir of our shock solutions, we could construct
other shock solutions that may match with the quasar spectra better.
By adjusting locations of the second shock, it is possible to match
with various asymptotic envelope solutions.

5. Supernova (SN) explosions and evolution of SN remnants. During
the late evolution stage of a massive star, the photodisintegration
and the neutronisation of the iron core will reduce the pressure of
the inner stellar core, triggering off a rapid core collapse. When
the inner core stops collapsing and rebounds the infalling materials,
shocks will form and propagate outwards through the inner core,
stellar interior and envelope. As the SN envelope expands during
the explosion of a SN, it will interact with the circumstellar
medium (CSM) and drive another shock through the powerful stellar wind.
We note that the speed of SN ejecta could be ultra-relativistic (e.g.,
Blandford \& McKee 1976; Ostriker \& McKee 1995; Bisnovatyi-Kogan
\& Silich 1995) and a gas may be modelled as polytropic (e.g.,
Goldreich \& Weber 1980). In this sense, our isothermal shock
solutions may not be directly applicable. However, the relevant
concepts and a variety of possibilities shown in this paper are
expected to be work for a polytropic gas as well. For example, a
shock can not only match with a static envelope but can also match
with a wind, as the CSM is contributed to stellar winds in the late
evolution stage of a massive star.

By discussions above, we show that similarity shock solutions are
potentially applicable to various astrophysical systems, ranging
from stellar to galactic scales. There exist collapses, outflows
or both collapses and outflows with one or more outward moving
shocks in these systems.

\section{Summary and Conclusion}

In this paper, we explore the self-similar isothermal shock
solutions, in reference to the earlier work of Tsai \& Hsu
(1995), Shu et al. (2002) and Shen \& Lou (2004).

First, we have significantly expanded the Class I and Class II shock
solutions of Tsai \& Hsu (1995) to two families of infinitely many
discrete shock solutions matched with a static SIS envelope by
systematically searching for intersections in the $\alpha -v$ phase
diagram. These shock solutions have subsonic radial oscillations
(Shen \& Lou 2004). As the value of reduced central mass $m_0$
decreases, the number of stagnation points in the corresponding $v(x)$
solution increases. For a given $m_0$, the location $x_{\ast}$ of
the solution crossing the sonic critical line is fixed. The shock in
the solution with two stagnation points is an accretion shock. It is
also possible to construct various similarity shock solutions with
different values $m_0$ to match with asymptotic breeze solutions.

Secondly, we further obtained shock solutions with twin shocks matched
with a static SIS envelope. This type of shock solutions contains an
inner accretion shock and an outer expansion shock. Both shocks travel
outward with constant yet different speeds. The travel speed of the
accretion shock is about 0.6 times the sound speed and that of the
expansion shock is about the sound speed. By adjusting the outer
expansion shock location, twin shock solutions
can match with various asymptotic solutions [e.g., inflow (accretion)
or outflow (wind)]. We apply this class of shock solutions to bridge
the AGB phase and pPN phase. By the property of EECC shock solutions,
we suggest to model accretion processes with shocks around black holes.

Finally, by allowing different yet constant temperatures on two
sides of a shock (i.e., $\tau>1$ with the downstream temperature
higher than the upstream temperature), we derived shock solutions
in a two-temperature fluid to match with various asymptotic
solutions at $x\rightarrow +\infty$.
These shock solutions may describe cloud collapse leading
to star formation and expansion of H{\sevenrm II} regions
surrounding massive OB stars.

In our model analysis, we pursue a simple isothermal scenario.
This is a special case of the more general polytropic treatment
(Cheng 1978; Goldreich \& Weber 1980; Yahil 1983; Bouquet et al.
1985; Suto \& Silk 1988; Maeda et al. 2002; Harada et al. 2003;
Fatuzzo et al. 2004; Wang \& Lou 2005 in preparation). Shock
solutions have also been derived in the polytropic case (Lou \&
Gao 2005, in preparation). Polytropic shock solutions should
have an even wider range of astrophysical applications.

\section*{Acknowledgments}
We thank the referee Dr. T. W. Hartquist for comments and
suggestions to improve the manuscript presentation. This research
was supported in part by the ASCI Center for Astrophysical
Thermonuclear Flashes at the University of Chicago under Department
of Energy contract B341495, by the Special Funds for Major State
Basic Science Research Projects of China, by the Tsinghua Center for
Astrophysics, by the Collaborative Research Fund from the National
Natural Science Foundation of China (NSFC) for Young Outstanding
Overseas Chinese Scholars (NSFC 10028306) at the National Astronomical
Observatories of China, Chinese Academy of Sciences, by NSFC grant
10373009 at the Tsinghua University, and by the Yangtze Endowment
from the Ministry of Education through the Tsinghua University. The
hospitality and support of the Mullard Space Science Laboratory at
University College London, U.K., of Astronomy and Physics Department
at University of St. Andrews, Scotland, U.K., and of Centre de
Physique des Particules de Marseille (CPPM/IN2P3/CNRS) et Universit\'e
de la M\'editerran\'ee Aix-Marseille II, France are also gratefully
acknowledged. Affiliated institutions of YQL share the contribution.

\appendix
\section[]{ Solution Matching }

\begin{figure}
\begin{center}
\includegraphics[scale=0.45]{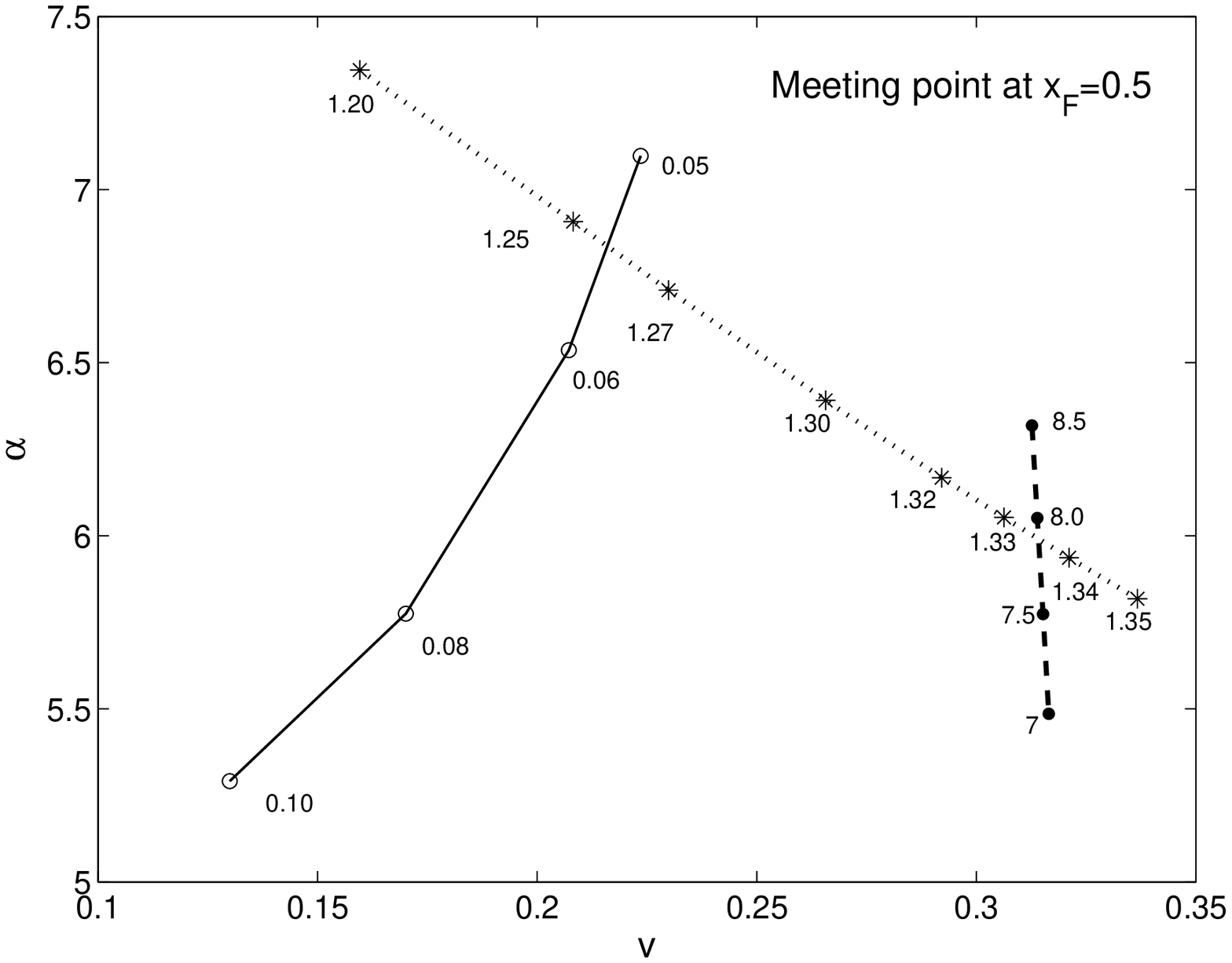}
\caption{The phase diagram of $v$ versus $\alpha$ at the meeting point
$x_{\rm F}=0.5$. Each asterisk symbol denotes an integration back from
the shock location $x_{s}$ with its value shown explicitly; each solid
dot denotes an integration from $x\rightarrow0^{+}$ for an inner LP
solution [equation (\ref{eq LP solution})] with the parameter $B$ marked
explicitly; each open circle denotes an integration from $x_{\ast}$ with
its value shown explicitly and this integration starts from a type 2
eigensolution across the sonic critical line $x-v=1$. For Class I shock
solution, we determine the location $x_{\ast}=0.0544$ where the flow
solution crosses the sonic critical line analytically and the shock
location $x_s=1.26$ from the intersection point of the solid and dotted
curves in the $v-\alpha$ phase diagram. For Class II shock solution, we
deterimine the reduced core density $B=7.9$ and the shock location at
$x_s=1.34$ from the intersection point of the dashed and dotted curves.}
\label{fig Tsai and Hsu match}
\end{center}
\end{figure}

We determine the parameters $x_{\ast}$, $x_s$ and $B$ by continuously
varying relevant parameters to search for intersections of the two
curves $v$ and $\alpha$ in the phase diagram of $v$ versus $\alpha$
where the meeting point is taken to be $x_{\rm F}=0.5$. We integrate
nonlinear ODEs (\ref{eq ODE1}) and (\ref{eq ODE2}) from
$x_{\ast}\ (<x_{\rm F})$ by using the type 2 eigensolutions across
the sonic critical line as specified by equation
(\ref{eq critical solution (2)}) to $x_{\rm F}=0.5$ for Class I
solutions or independently, from $x\rightarrow0^{+}$ by using the
LP-type solution specified by (\ref{eq LP solution}) as the starting
condition to $x_{\rm F}=0.5$ for Class II solutions. In this manner,
we obtain a pair of $v$ and $\alpha$ in the so-called phase diagram.
By gradually varying parameter $x_{\ast}$ (for Class I) or $B$
(Class II), the values of $v$ and $\alpha$ at the meeting point
$x_{\rm F}$ will change continuously in the $v-\alpha$ phase diagram.
In Fig.~\ref{fig Tsai and Hsu match}, we show such phase curves in
solid and dashed curves for Classes I and II solutions, respectively.

Meanwhile, we choose different shock locations and apply the
jump condition (\ref{eq shock condition (1) for tau=1}) to obtain
compatible $v_d$ and $\alpha_d$ as the starting condition and
integrate back from $x_s$ to $x_{\rm F}=0.5$ to determine values
of $v$ and $\alpha$ at $x_{\rm F}=0.5$; the dotted curve in
Fig.~\ref{fig Tsai and Hsu match} is obtained this way and
from the intersections between the dotted curve and the solid
and dashed curves, we determine parameters $x_{\ast}$, $x_s$
and $B$ for relevant similarity shock solutions.

\begin{figure}
\begin{center}
\includegraphics[scale=0.45]{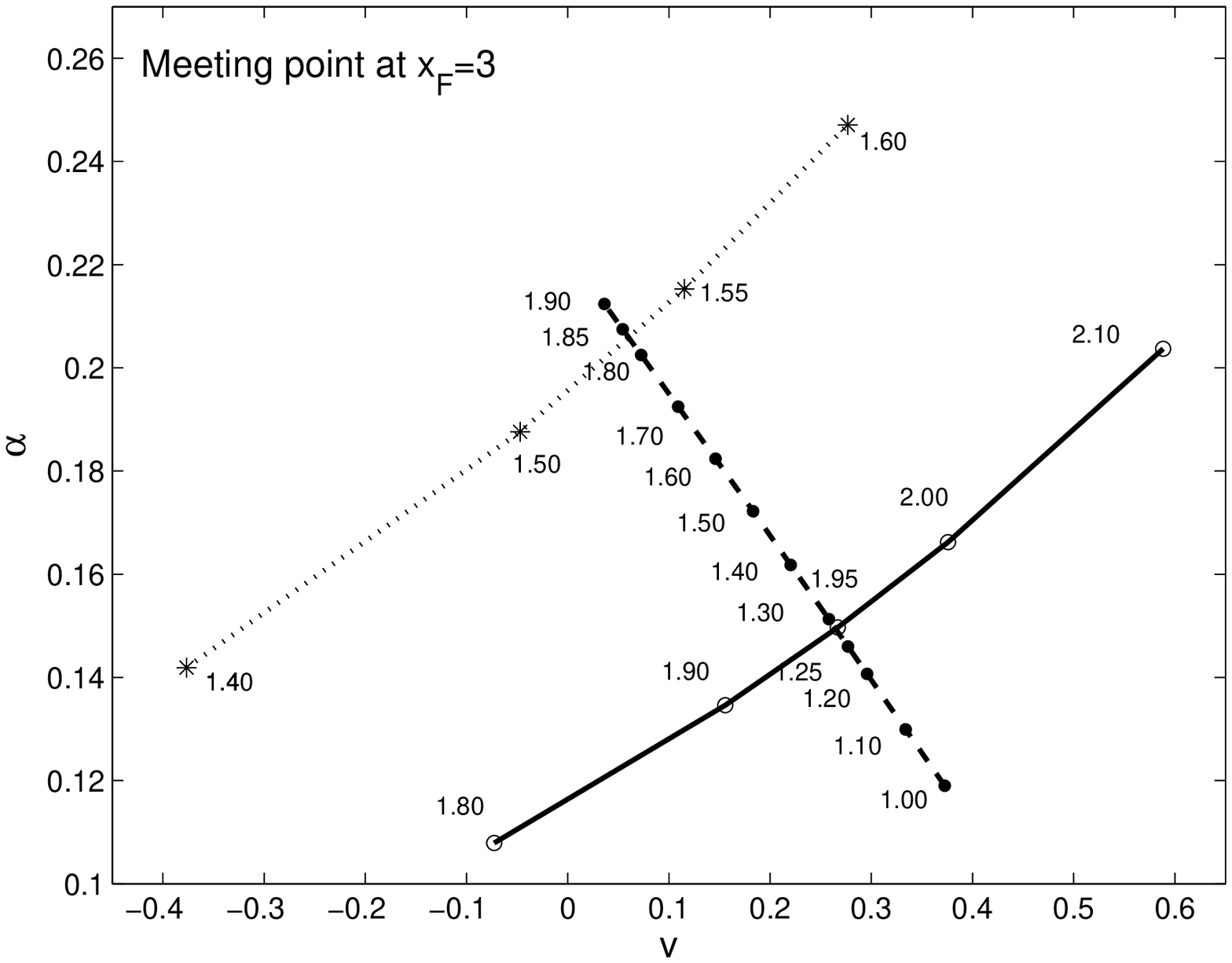}
\caption{The phase diagram of $v$ versus $\alpha$ at the meeting point
$x_{\rm F}=3$. Each asterisk symbol indicates an integration from the
shock location $x_{s}$ with its value shown explicitly for $B=4$;
each open circle symbol indicates an integration from the shock
location $x_{s}$ with its value shown explicitly for the case of $B=1$;
each solid dot denotes an integration from $x\rightarrow+\infty$
(practically, we take $x=20$) to use the analytical asymptotic solutions
(\ref{eq asymptotic solution x-infty}) with the velocity parameter $V=0$
and the mass density parameter $A$ with its value marked explicitly.
From the intersections of the dashed curve with the dotted curve and the
solid curve, respectively, we determine the shock location $x_s$ and the
value of mass density parameter $A$ for $B=4$ and $B=1$, respectively.}
\label{fig shu match}
\end{center}
\end{figure}

The mass parameter $A$ and shock location $x_s$ for the shock
solution of Shu et al. (2002) for a given $B$ can be determined
from the phase diagram of $v$ versus $\alpha$. Two specific
examples are shown in Fig.~\ref{fig shu match} for $B=1$ and
$B=4$, with the meeting point chosen at $x_{\rm F}=3$. The
numerical procedure is similar to that described in \S\ 3.1.1.

\bsp

\label{lastpage}

\end{document}